\documentclass[10pt,sigconf,letterpaper,nonacm,natbib=false]{acmart}

\usepackage[subtle]{savetrees}
\usepackage{setspace}
\usepackage[margin=0pt,skip=1pt,font={stretch=0.9}]{caption}

\renewcommand{\sfdefault}{cmss}

\usepackage{booktabs} 
\usepackage{outlines}
\usepackage[maxbibnames=99, backend=biber, bibencoding=utf8, abbreviate=true, %
dateabbrev=true, isbn=false, doi=false,]{biblatex}
\AtEveryBibitem{\clearlist{location}%
  \clearfield{numpages}%
  \iffieldundef{eprint}{}{\clearfield{url}}%
  \ifentrytype{misc}%
      {}%
      {\clearfield{url}}%
  \ifentrytype{inproceedings}%
    {\iffieldundef{series}{}{
      \savefield{series}{\seriesname}%
      \clearfield{series}%
      \restorefield{booktitle}{\seriesname}%
    }}%
    {}%
}
\usepackage{bytefield}

\setcopyright{none}
\renewcommand\footnotetextcopyrightpermission[1]{}

\usepackage{url}
\usepackage{calc,multirow}
\usepackage{latexsym,amsmath}
\usepackage{array}
\newcolumntype{M}[1]{>{\centering\arraybackslash}m{#1}}

\usepackage{xspace}
\usepackage{xcolor}
\usepackage{algorithm,algpseudocode,xcolor}
\usepackage[shortlabels]{enumitem}
\usepackage{tikz}
\usepackage{svg}

\usepackage{subcaption}
\usepackage[binary-units]{siunitx}
\usepackage{listings}
\AtBeginDocument{%
  \DeclareSIUnit\bit{b}%
}
\usepackage{nth}

\newcommand{\headline}[1]{\vspace{+0.05in}\noindent\textbf{#1}}

\setlist[itemize,1]{leftmargin=\dimexpr 4mm}

\begin{document}

\title{\huge Implementing packet trimming support in hardware}

\author{
  Adrian Popa$^*$,
  Dragos Dumitrescu$^{* \ddagger }$,
  Mark Handley$^{\circ *}$,\\
  Georgios Nikolaidis$^\triangle$,
  Jeongkeun Lee$^\triangle$,
  Costin Raiciu$^{* \ddagger }$ \\
  \em                                                                                                                                       
  $^*$ Correct Networks,                                                                                                                   
  $^\circ$ UCL,
  $^\triangle$ Intel,
  $^\ddagger$ University Politehnica of Bucharest}


\renewcommand{\shortauthors}{Popa et al.}

\begin{abstract}
  Packet trimming is a primitive that has been proposed for datacenter
networks: to minimize latency, switches run small queues; when the
queue overflows, rather than dropping packets the switch trims off the
packet payload and either forwards the header to the destination or
back to the source.  In this way a low latency network that is largely
lossless for metadata can be built.

Ideally, trimming would be implemented as a primitive in switch ASICs,
but hardware development cycles are slow, costly, and require
demonstrated customer demand.  In this paper we investigate how
trimming can be implemented in existing programmable switches which
were not designed with trimming in mind, with a particular focus on a
P4 implementation on the Tofino switch ASIC.  We show that it is
indeed possible to closely approximate idealized trimming and
demonstrate that trimming can be integrated into a production-grade
datacenter switch software stack.

\end{abstract}

\maketitle

\section{Introduction}

Congestion control algorithms rely on network feedback to adjust the
sending rate to match and adequately share available capacity. While
in the Internet such feedback is almost always packet-loss, datacenter
networks enable the deployment of a wider range of congestion feedback
including ECN marking and link-layer priority flow control for
intra-datacenter traffic, as each datacenter is managed by a single
organisation.

The available type of congestion feedback, coupled with buffer sizing,
has a direct effect on the performance of congestion control
algorithms. In datacenters, loss-based congestion controllers need
very large buffers (megabytes) to cope with bursty many-to-one traffic (incast),
or they suffer congestion collapse \cite{minrto,dctcp}; large (shared)
buffers however add latency to innocent traffic on the same
egress port. ECN marking helps reduce buffer usage for long-running flows,
while still needing large buffers to absorb incast bursts
\cite{dctcp}. Priority Flow Control eliminates congestion-related packet-loss so it is
apparently ideal for incast traffic, but it can delay innocent traffic
and result in poor throughput; it is often used with large buffers and
in conjunction with ECN, to avoid triggering frequently \cite{dcqcn}.

In summary, the congestion feedback mechanisms available in production
today force congestion controllers to tiptoe around a tradeoff between
utilization, latency and loss on one hand, and between solving
particular problematic traffic patterns such as incast and causing
damage to bystander traffic crossing the same switches. Recent
research papers have however shown that, with the network hardware
support for queue overload, it is possible to build stacks that
achieve high utilisation, have almost perfect incast behaviour without
hurting innocent traffic all the while using small packet buffers
\cite{ndp,cp,bts,homa,aeolus,eqds}.

One very promising overload behaviour is packet trimming, where switches
send control information (in the form of packet headers) to the
destination or the source of the traffic when an output queue is full
\cite{ndp,cp,bts}: instead of dropping or marking the packet, trimming
removes the payload and only forwards the header, typically with
priority, to the destination or the source, depending on the scheme chosen.
The benefit of trimming is that it allows very small per port buffers to be
used (10-20 packets \cite{ndp,eqds}) while ensuring packet headers are not lost so hosts know
unambiguously what happened to a packet. Trimming enables the creation of host stacks that are near-optimal for
all traffic patterns, including incast and permutation, effectively turning
datacenter networks into an approximation of a perfect non-blocking switch
connecting all hosts \cite{ndp,eqds}.

Can endhost stacks use trimming in real datacenter networks? Existing
works provided NetFPGA or software switch implementations
\cite{ndp,cp}, but these are neither fast nor stable enough for use in
production networks. An ASIC implementation would be ideal, together
with the control hooks, but hardware development cycles are slow
(years), costly and typically require strong interest from potential
customers before any investment from hardware vendors. A quicker and indeed
workable deployment approach is to use existing switch functionality,
either programmable or fixed-function, to implement trimming or
approximations of it.

In this paper we analyze to what extent trimming can be supported on production
switches today. We find that, surprisingly, many switches being used in production do have the building
blocks needed to support packet trimming, namely the ability to mirror/deflect packets
about to be dropped and the ability to only copy the header of the dropped packet; this header can
then be enqueued in a higher priority queue on the same output port, typically after being recirculated
through the switch. Such functionality is broadly available in production switches; it is used by operators to
help monitor packet losses in their networks by aggregating drop information to a centralized monitor.

Mirror-on-drop is thus a generic approach that can implement trimming
on a variety of switch architectures; this strategy, however, does
have some limitations which stem from the architecture of the
switches; these limitations can make the hardware implementation of
trimming significantly differ from the ideal behaviour. We analyze these
limitations in simulation, and then propose a concrete design for
Intel's Tofino switching ASIC that provides trimming support. Our evaluation
shows that it is possible to build an almost ideal implemementation of
trimming in hardware.

\section{Packet trimming overview}

The concept of packet trimming was first suggested as cut
payload\cite{cp}, where instead of dropping a packet a switch
simply cuts the payload and forwards only the packet header.  NDP \cite{ndp}
refined this concept to avoid potential livelock situations and to
provide early negative acknowledgments, allowing a trimmed packet to
be retransmitted sufficiently quickly that when a switch queue only
just overflows due to a small burst, the resent packet can reach the
queue before it has had time to drain, ensuring that the trimming has
no negative impact.  NDP does this by queuing the trimmed packet header in a
higher priority queue than untrimmed data packets.

Ideally, strict priority queuing is not used with NDP, but rather a byte-based weighted fair
queuing regime should be used instead.  This ensures that some data
packets are still forwarded when many packets are being trimmed, but
all the early trimmed headers will jump the queue of data packets.  This provides
improved performance and eliminates livelock in extreme circumstances.

Trimming can still lead to header drops in extreme circumstances when the per port
header buffer is already full; this happens in incasts with thousands of senders. In such
cases, the NDP paper proposed that the header should be returned to the sender instead,
to avoid dropping it.

Recently, the header return-to-sender approach has been proposed as a
standalone solution called Source Flow Control (SFC) \cite{bts}, where
the trimmed headers start to be returned to the sender when the queue
passes a threshold, rather than only when it cannot be forwarded to the
destination.  SFC is undergoing standardization at the IEEE.

In summary, packet trimming is a switch-based feature that behaves as follows:
\begin{enumerate}
\item When an output port buffer is full (NDP, SFC) or when some watermark has been hit (CP, SFC),
  trim the packet to its header instead of dropping it,
\item
  Optionally set some bits in the trimmed header (i.e. trimmed bit in NDP).
\item Optionally reverse
  the source and destination addresses for return-to-sender operation (SFC, NDP when
  the header queue is full).
\item Enqueue the resulting header either in (a) the same output queue
  (CP); (b) a higher priority queue (NDP); (c) the output queue destined for the original sender (SFC, NDP when
  the header queue is full).
\end{enumerate}

\subsection{Benefits of packet trimming}

Trimming allows very small data packet queues (as low as ten MTU-sized
packets \cite{ndp,eqds}) to be used in datacenter switches,
guaranteeing that packets cannot be queued within the network for any
significant amount of time.  As packets are almost never dropped
entirely, the receiver gains a complete view of the instantaneous
demand from all senders (when the headers are sent forward) and the
senders very quickly receive feedback from receivers or the switches
as to a packet's fate, avoiding the need for conservative
retransmission timers to prevent congestion collapse.  Such
conservative timers typically have a detrimental affect on performance \cite{minrto}.

As packet headers are generally not lost, trimming allows higher-layer
protocols to use per-packet ECMP\cite{spray} (packet spraying) without the risk of
confusing reordering with packet loss.  This allows networks using
Clos or similar redundant topologies to be load-balanced much more
effectively, avoiding flow collisions which have been shown to reduce
throughput in the worst case by as much as 60\% \cite{hedera,mptcp-dc}.

Trimming is especially effective with incast traffic patterns because
it does not require a tradeoff between queue latency and packet
loss. Switch buffers can be kept very small yet there is (almost) no
packet loss with incast; instead only the packet payload is dropped,
allowing senders to know the fate of their packets almost immediately.

Prior works have shown that trimming enables near-perfect network
utilization , incast behaviour and short flow completion times, at
scale \cite{ndp,eqds}, providing a big improvement over the status quo
represented by DCTCP \cite{dctcp}, DCQCN \cite{dcqcn}, MPTCP
\cite{mptcp-dc}.

Trimming, then, is a very effective basic building block for
datacenter networks.  Unfortunately current network switches were not
designed to support trimming, which leaves protocol designers with a
chicken-and-egg problem: they cannot design protocols to assume short
queues and use per-packet ECMP because they cannot assume trimming
will be available; at the same time switch vendors have no demand from
customers for trimming support because no protocols could take
advantage.

\begin{figure}
  \center
  \includegraphics[width=1.0\columnwidth]{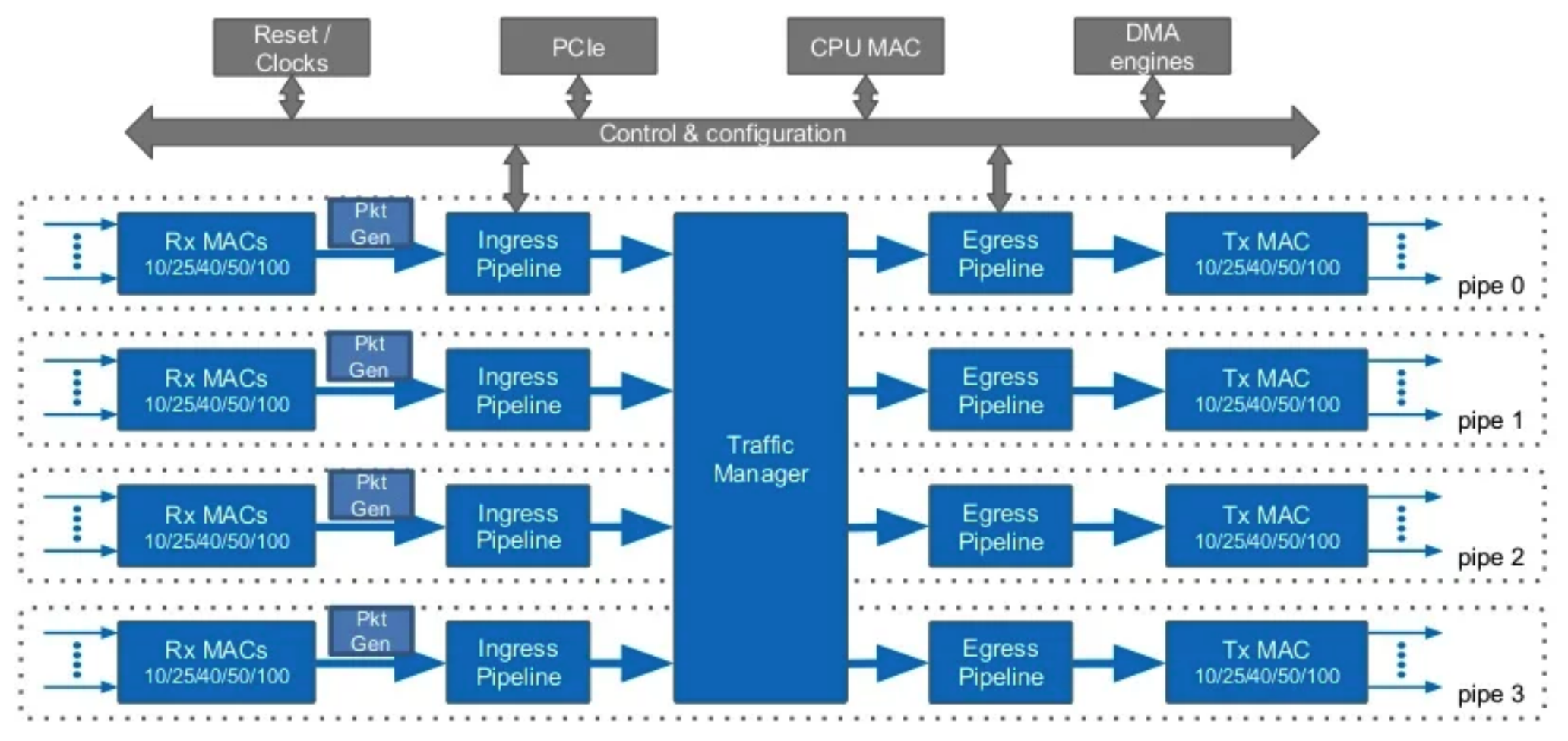}
  \caption{Simplified Tofino block diagram.}
  \label{fig:tofino-arch}
\end{figure}

Fortunately this cycle can be broken today by implementing trimming
support in current programmable switches, but doing so is not
completely trivial.

\section{Implementing trimming in programmable switches}
\label{sec:trimming}

Trimming is conceptually a function of an output-queued switch where
trimming is done when attempting to enqueue and the queue is either
full or has reached a predefined watermark.  To emulate this in a
programmable switch, we need to consider the switch architecture. The
building blocks at our disposal may differ somewhat between different
programmable switches.  For example, a Tofino switch consists of
four programmable ingress pipelines, each fed from sixteen 100 Gb/s
ingress ports, a central traffic manager which implements queuing and
interconnection between pipelines, and then four programmable egress
pipelines each feeding sixteen 100Gb/s egress ports, as shown in
Figure~\ref{fig:tofino-arch}.  Note that trimming cannot be
implemented in the egress pipelines leading to the egress port because
these pipelines are downstream of the queues; it should ideally be
implemented in the traffic manager (not programmable) or in the ingress
pipeline where queue information is lacking.

\headline{Deflect-on-Drop.}  Many switches feature a form of {\em
deflect-on-drop} (DoD) functionality in their equivalent of the
traffic manager.  When the queue for an egress port fills, the traffic
manager can be configured to deflect the packet to another port
instead of dropping it.  DoD is implemented primarily to enable
network operators to analyze the packets that were dropped, and is
supported by most switches, not just programmable ones.  Our strategy
is to re-use this functionality to deflect packets that would have
been dropped to an internal port used for recirculation.  We refer to
this internal port as the DoD port.  How can we use this for trimming?

Intel Tofino and Tofino 2 switching ASICs support deflect-on-drop, but 
deflected packets carry their full payload and not just the headers.
Tofino-based switches have internal recirculation ports for each pipeline
which we can use to forward deflected traffic. This recirculated traffic can then
be trimmed to the desired length using mirroring. On Tofino, for
example, this means that we can recirculate up to 100Gb/s of
untrimmed packets using one dedicated internal port per pipeline.
If all ports on the same ingress pipeline start
sending to a single egress port, that port will be oversubscribed
16:1.  1.5Tb/s of traffic would then be deflected to the same DoD
port, resulting in a large queue followed shortly by large packet
loss.  Thus in the Tofino family of switches, we need to prevent such
large overloads of DoD capacity from happening in the first place.

Some switches including the Broadcom Trident~4 and NVidia Spectrum~2
additionally support the ability to configure their equivalent of the traffic
manager to {\em mirror just the packet header} into a DoD queue.
Once there, the header can be recirculated and then queued in a priority
header queue for the original destination port (NDP) or the
port back to the sender (SFC). In such a switch, trimming largely happens in the traffic
manager with any additional processing such as setting a trimmed bit
(NDP) or swapping addresses (SFC) happening in the DoD pipeline before
recirculation. One downside of such an implementation is that
trimmed packets must all be recirculated, so they will be delayed by a
few packets relative to non-trimmed packets. This is overall a good
approximation but as we will see in Subsection
~\ref{subsec:basic_mechanism_comparison}, we can
take advantage of Tofino's programmability to improve performance even further.

\headline{Trimming in ingress.}
If the ingress pipelines had access to the instantaneous queue size
for traffic from all ingress ports to the respective egress port, then
trimming might be performed in P4 in the ingress pipelines.  In Tofino
2, such queue length information is available at registers in the ingress
pipelines, which are updated by a thread separate to the normal
switching control with minimal feedback delay compared
to recirculation~\cite{q_stats}.

With queue size information available, when an incast occurs and the
queue size reaches the target threshold, trimming can begin in the
ingress pipeline.  However a P4 program performing trimming must read
the queue size in an early stage to decide a packet's fate.  When
trimming starts, there can already be multiple packets in the later
stages of the ingress pipelines that have not yet been enqueued and are
not account for in the queue size information.  For example, if there are four pipeline stages downstream
of the stage reading the queue size, and eight pipelines in our switch,
there may be at most 32 untrimmed packets that will hit the output
queue after the switch starts trimming.  Most of these packets will
then be deflected to the recirculation ports.  On Tofino 2, we can use
one recirculation port for each eight ingress ports, so the DoD queues will grow
to about 4 packets each, before dropping back after ingress-trimmed
packets start reaching the output queue.  Compared to a
mirror-on-drop solution, most trimmed packets on such a Tofino 2
implementation will be ingress trimmed and so see lower latency, but
some at the start of an incast that traverse DoD will see
higher latency due to the build-up of the DoD queue.


The situation is more complicated on Tofino, where queue size
information is unavailable to the ingress pipeline.  We can still
perform trimming, but we need to be a little smarter in how we go
about doing it.  We can keep track of all the traffic destined for
each output port using software in the ingress pipeline.  To do this
we use a {\em meter} to emulate a virtual output queue for each egress
port; when that virtual queue fills, we trim the packet before it goes
to the traffic manager for buffering.  This solution eliminates the
lag seen when reading the real queue size, so packets trimmed in this
way see the lowest latency of all the solutions.

Unfortunately though, there is a downside: meter information cannot be
shared between ingress pipelines on Tofino.  The fact that information
is local to a pipeline is not fundamental in programmable switches,
but to perform at very high speeds it is necessary to parallelize
hardware, and any shared state between parallel building blocks is a
potential bottleneck.  What are the implications of this, and what can
we do despite this constraint?

Suppose we implement meter-based trimming independently on each
ingress pipeline.  If all the traffic for an output port came from a
single ingress pipeline, all trimming would occur in the ingress
pipeline, perfectly matching our output-queued trimming ideal.  The
problem arises when traffic arrives via multiple ingress pipelines
destined for one output port.  Each ingress pipeline will trim
assuming it is the only source of traffic, but the resultant trimmed
packet stream will overflow the egress port queue in the traffic
manager.  These excess packets will then be deflected to the DoD
ports.  With meter-based ingress trimming reducing the overload, is
there enough DoD capacity to cope?

On Tofino, we can use one recirculation port for each of four ingress
pipelines, and each pipeline has sixteen ingress ports.  One might
think that the worst-case traffic pattern for the Tofino DoD pipeline
would be when 100Gb/s arrives from 63 input ports destined for a
single output port.  In this case each ingress pipeline would use its
local meter to trim down to 100Gb/s, resulting in 400Gb/s destined for
the output port.  100Gb/s will be forwarded directly, leaving 300Gb/s
to be redirected to the four recirculation ports on those pipelines.
In this case there is enough capacity, and all the remaining trimming
is performed there.

\begin{figure}
  \center
  \includegraphics[width=1.0\columnwidth]{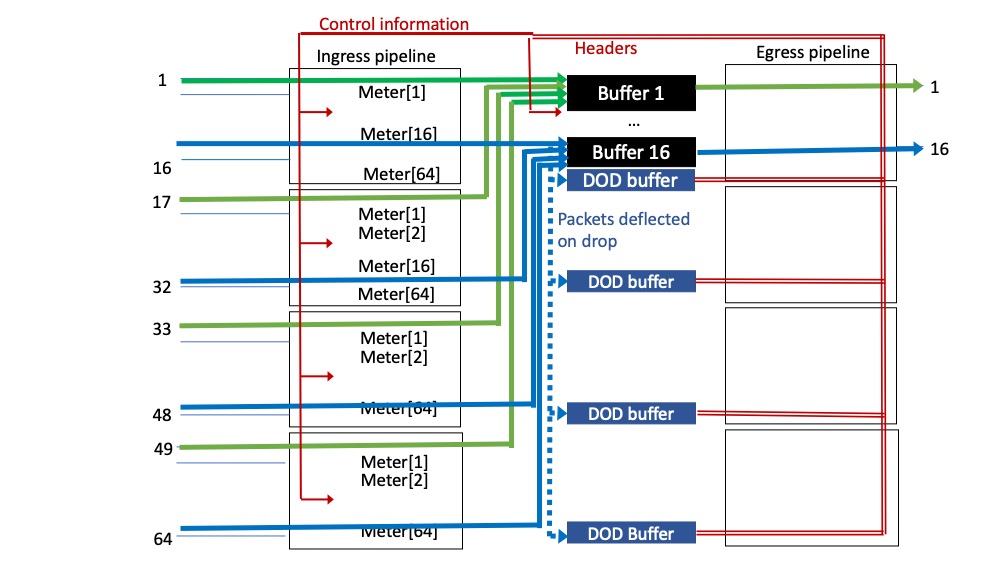}
  \caption{Worst-case DoD Scenarion for Tofino}
  \label{fig:worst_case_dod}
\end{figure}

In practice such a 63:1 incast is not the worst-case scenario.
Consider what happens when 100Gb/s of traffic arrives on port $p$ on
every pipeline destined for port $(p+1)\%16$ on egress pipeline 0.
For example, traffic arrives on ports 0, 16, 32 and 48 all destined
for port 1; simultaneously traffic arrives on ports 1,17,33 and 49
destined for port 2 and so on, up to traffic arriving on ports
15,31,47 and 63 destined for port 0.  Effectively sixteen parallel 4:1
incasts occur simultaneously, as shown in
Figure~\ref{fig:worst_case_dod}.

In this case, from the point of view of a single ingress pipeline,
only 100Gb/s of traffic arrives for each output port, so no trimming
is performed in the ingress pipeline.  Now, 400Gb/s of traffic arrives
at the traffic manager for each of the 16 output ports targeted.
1.6Tb/s can be forwarded directly, leaving 4.8Tb/s to be diverted to the
four recirculation ports, which will be oversubscribed 12x,
 with their queues rapidly growing, leading to high packet loss.

The challenge of implementing trimming is to handle this 16 x 4:1
incast scenario while simultaneously being as close to optimal as
possible in scenarios between this and the 63:1 incast scenario which
itself requires no special mechanisms beyond ingress trimming and
trimming in the DoD pipeline.  To do this requires a control loop that
adjusts the trimming rate in the ingress pipelines.

\section{Trimming Control Loop Design}
\label{sec:solution}

DoD is very effective for coping with small transient overloads, but
if the queue feeding the recirculation pipeline starts to grow, then
DoD itself can become a bottleneck.  In such circumstances we require
ingress pipeline trimming to become more aggressive.

When the ingress pipeline performs trimming based only on local
traffic meters, trimming traffic to match an egress port's link speed,
we refer to this as {\em optimistic mode} as the ingress pipeline is
assuming no other traffic for the port exists from other pipelines.
When this is not the case, to increase ingress pipeline trimming a
congestion signal can be recirculated from the egress pipeline when
DoD packets arrive there.  This congestion signal should trigger
increased packet trimming in the ingress pipeline to prevent the
recirculation queue growing excessively.




When evaluating the possible designs our goals are to minimize the
number of packets trimmed in excess of those trimmed by an ideal
implementation, to ensure quick feedback by keeping recirculation
queue sizes small, and to ensure fairness across flows from different
pipelines.

\vspace{2mm}\noindent {\textbf{Hard vs soft state.}}
Once a congestion signal from the recirculation port enables more
aggressive trimming, we need a way to go back to optimistic mode when
traffic abates and the recirculation queue drains.  In principle we
could adopt a hard-state approach, sending an ``all clear'' signal
from the recirculation port, but this would have to be triggered from
a timer or directly from the recirculation queue becoming empty, which
would be complex to implement.

The alternative is to adopt a soft-state approach where the congestion
signal triggers the congestion response for a predefined amount of
time or a predefined number of packets before reverting to regular
optimistic mode.  

There are two soft-state approaches we could take:
\begin{itemize}
\item Each congestion packet instructs the pipeline to trim a \textit{fixed
  number of packets}; the pipeline goes into pessimistic mode until this number
  of packets is received and trimmed.

\item Each congestion packet instructs the pipeline to go into
  pessimistic mode for a \textit{fixed period of time}, starting from the
  receipt of the packet.
\end{itemize}

\vspace{2mm}\noindent\textbf{Trim N Packets.}
The simplest soft-state mechanism is whenever a packet leaves the
recirculation queue of pipeline A, instruct the ingress pipeline of A
to trim one packet destined for the same port.  Unfortunately the
recirculation port only sees packets that make it through the
recirculation queue, and packets can be entering this queue much
faster than it can drain - 12x faster in the worst case.  To handle
the worst case then, each packet making it into the recirculation
queue would have to trigger a congestion response trimming at least 12 packets.

Trim N is quite simple to implement but is suboptimal: when a large
short-duration incast occurs and the recirculaton queue fills, as
this queue drains it will result in many congestion notification
packets being sent to the ingress pipeline, each instructing it to
trim at least 12 packets; if $n$ congestion packets are sent, the
pipeline will trim $12n$ packets.

If the incast is short lived, as would be expected if an end-to-end
control loop such as NDP is running, there will be no more incast
traffic to trim, and instead these trims will affect other packets
an arbitrary time later, reducing goodput and creating an
unnecessary coupling between the incast and subsequent unrelated
traffic.

\vspace{2mm}\noindent\textbf{Trim for a time period.}
The alternative soft-state mechanism we actually use is for a
congestion signal to indicate to trim traffic destined for the
specified destination port, starting now, lasting for $t$
microseconds.  This has two advantages: multiple signals arriving
right after one another don't greatly change the outcome, and the
effect of a short incast is inherently short.

Trimming all packets for duration $t$ would entail a form of
pulse-width modulation.  In practice though, with four ingress
pipelines on Tofino, we do not need to control to arbitrarily low
rates - on a switch with four pipelines and 100Gb/s ports the full control range only
needs to operate from 25Gb/s to 100Gb/s per output port per pipeline.

The simplest algorithm then would be for the ingress pipeline to trim
traffic to 25Gb/s for $t$ microseconds on receipt of a congestion
signal.  We refer to this 25Gb/s target rate as {\em pessimistic
  mode}. This algorithm does work, but we found that the trimming
period $t$ needs to be fairly long to be stable in all circumstances,
resulting in some degree of underutilization.

To smooth the rate increase as congestion abates we added a third
ingress pipeline trimming mode, so now we have:
\begin{itemize}
\item {\bf Optmistic mode}: trim to 100Gb/s
\item {\bf Half-pessimistic mode}: trim to 50Gb/s
\item {\bf Pessimistic mode}: trim to 25Gb/s
\end{itemize}
We use three fixed-rate meters in the P4 ingress pipeline to implement
these modes and switch between them based on congestion signals from
the recirculation port.  Each ingress pipeline remembers which mode it is operating in for each
output port.  One packet arrival, the
corresponding meter is used to decide whether or not to trim the
packet.  When a recirculated congestion signal arrives, the
corresponding port is set to pessimistic mode for a predetermined
period of time (6us -- long enough to trim 12 packets).  When this time
period expires the port switches to half-pessimistic mode for a
further 6us, before returning to optimistic mode if no furthger
congestio signal is received.

\vspace{2mm}\noindent\textbf{Send signal to all pipelines.}
A final design decision is whether the congestion signal should be
sent to one pipeline or all of them.  The obvious choice is to send
the congestion signal only to the pipeline which originated the
deflected packet.  It turns out that this adds delay to the control loop
- the signal takes longer to reach all ingress pipelines - and this
necessitates running larger time constants.

We also found that send-to-one can results in unfairness and potential oscillations
between pipelines. Consider the following packets in the DoD queue,
where a packet is represented by its intended egress port and the front of the queue is at the left:
\begin{description}
\item{Pipe 1 DoD:} 1,1,1,1,1,1,1,1,2
\item{Pipe 2 DoD:} 2,2,2,2,2,2,2,2,1
\end{description}

Pipe~1 will go into congestion mode for Port~1 for 8+$t$ time slots,
and Pipe~2 will do the same for Port~2. During this time, traffic for Port~1
from Pipe~2 can run at line rate, taking~80\% of the egress
capacity and filling the DoD queue with 1’s. Traffic on Port~2
arriving from Pipe~1 call also run at line rate. 
Beyond the temporary unfairness, such oscillations 
can continue, preventing the recirculation queues from draining.

As a result we send all congestion signals to all pipelines; this
ensures rapid responses across all pipes to queue increases.

\section{Tofino implementation}
\label{sec:tofinoimpl}

We have implemented two versions of on Tofino: \textsf{ndp.p4} is a 
standalone IPv4 router with trimming support,
the second implementation is fully integrated with \texttt{switch.p4}, a production-grade
datacenter switch with SONiC support. \textsf{ndp.p4} is more
self-contained so we will focus on that first.

\headline{ndp.p4}. This implementation contains a
longest-prefix match routing table, an ARP table and a rewriter in
egress, plus extensions to support trimming which are mostly in the
ingress pipeline.

\begin{figure}
	\center
	\includegraphics[width=0.9\columnwidth]{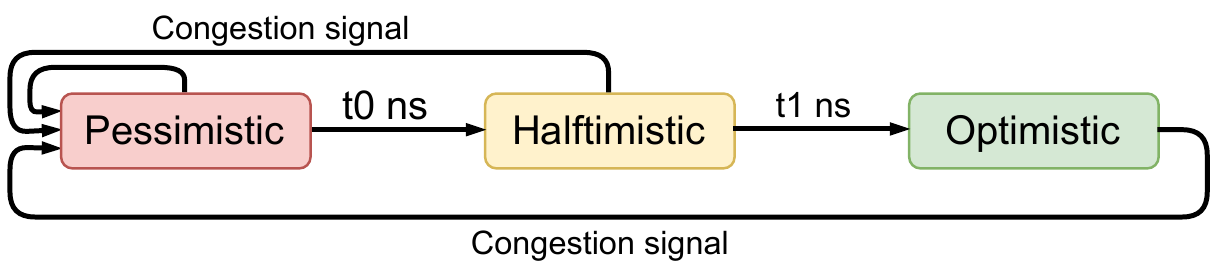}
	\caption{Port trimming state machine}
	\label{fig:port-states}
\end{figure}

When a packet overflows the port queue, is deflected, and is detected
in egress, a notification packet is sent to all pipelines in ingress
on a special recirculation port. When this notification arrives in
ingress, the output port which experienced the loss transitions into
pessimistic state, as shown in figure \ref{fig:port-states}.  In the
absense of congestion notification, the port will transition from
pessimistic state to half-pessimistic (aka halftimistic) state after
time $t_0$ elapses, and then back to optimistic after $t_1$.  Any
congestion notification causes a transition to pessimistic and restarts
the $t_0$ timeout.
%
%

In the ingress pipeline we maintain three stateful meters for each egress port:
\textit{meter\_optimistic}, \textit{meter\_pessimistic} and
\textit{meter\_halftimistic}.  These are used in the corresponding
states and meter at line rate, 25\% of line rate, and 50\% of line
rate respectively.
 We also maintain two registers, \textit{t0\_reg} and
\textit{t1\_reg}, which are timestamps for state expiry, as in
Figure~\ref{fig:port-states}.

The code executed on every ingress packet is sketched out in this pseudo-code:

\lstset{
  basicstyle={\Small\ttfamily},
  stringstyle=\ttfamily,
  frame=single,
  emph={ON, if, else},
  emphstyle=\color{red},
  emph={[2]t0_reg, t1_reg, opticolor, pessicolor, halfcolor, color},
  emphstyle={[2]\color{blue}},
  }
\begin{lstlisting}
ON DoD got on port egress_port:
  t0_reg[egress_port] = now() + T0
  t1_reg[egress_port] = now() + T1
ON non-DOD packet:
  opticolor=meter_optimistic.execute(egress_port)
  pessicolor=meter_pessimistic.execute(egress_port)
  halfcolor=meter_halftimistic.execute(egress_port)
  color = opticolor
  if now() <= t0_reg[egress_port]:
    color = pessicolor
  else if now() <= t1_reg[egress_port]:
    color = halfcolor
  else:
    color = opticolor
    t0_reg[egress_port] = 0
    t1_reg[egress_port] = 0
  if color == RED:
    trim_packet()
\end{lstlisting}

%
%

\begin{figure}
	\center
	\includegraphics[width=0.9\columnwidth]{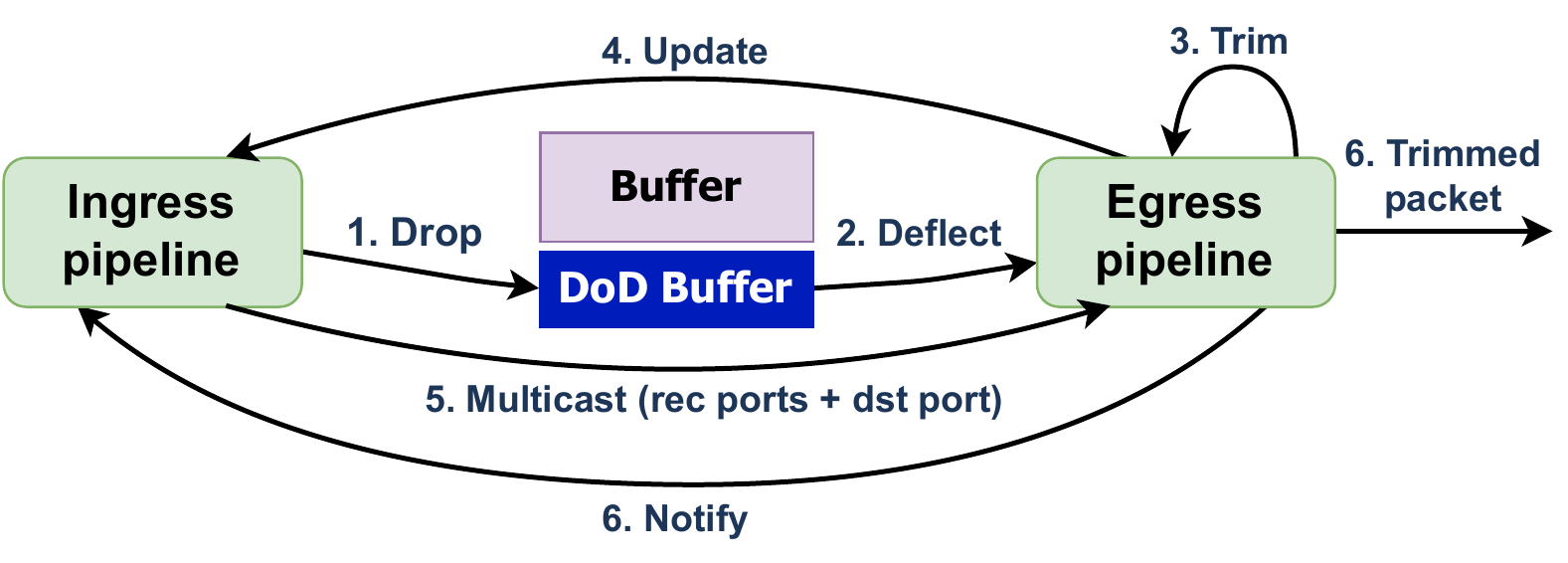}
	\caption{Congestion signaling}
	\label{fig:congestion-signalings}
\end{figure}

\headline{Congestion signaling}
To send the congestion signal to all ingress pipelines, we use control
messages which are recirculated (figure \ref{fig:congestion-signalings}).
First, when a deflected packet arrives in egress, we convert it into a control message
by trimming its payload and adding a new header at the beginning of the packet.
This new header is used to store information about the original packet and what type
of control message it is.

To trim the data packet we use one mirroring session for each pipe.  Each sets that
pipe's recirculation port as destination and instructs
the traffic manager to trim the packet. The newly
trimmed packet forms an UPDATE control packet which
will be sent back to ingress in that pipe only.

In the ingress parser we distinguish data packets from UPDATEs
by looking at the ingress port. When a packet arrives in ingress
from the recirculation port, we know that it is an UPDATE and 
look for the UPDATE header.

When the ingress pipeline receives an UPDATE packet,
it uses a multicast group to propagate the congestion
signal to all pipes and changes the packet type to NOTIFY.
The multicast group contains the recirculation ports of all pipes.
NOTIFY packets arriving in egress are recirculated to their respective
ingress ports; they instruct each pipe to transition the congested port
into \textit{pessimistic} state.

\headline{\texttt{switch.p4} integration.} The standalone \texttt{ndp.p4}
is good for testing and evaluation, but lacks many features of a production-grade
router.  For operational deployments we integrated \texttt{ndp.p4} into
\texttt{switch.p4}, a production-grade datacenter router implementation fully
integrated with SAI and SONiC.

We cannot quite insert the standalone implementation of NDP directly into \texttt{switch.p4} for several reasons:
\begin{enumerate}
	\item \texttt{switch.p4} takes up almost all stages of the Tofino. Fitting
	extra functionality is challenging.
	\item \texttt{switch.p4} employs custom signaling between ingress and egress
	pipelines to trigger specific behaviors (bridging information,
	dataplane telemetry, etc.). Harmonizing the notification mechanisms
	of \texttt{ndp.p4} with existing ones required
	careful re-design.
	\item SONiC, the control-plane of \texttt{switch.p4} is a very complex
	piece of software; making \texttt{NDP} fit in this picture is difficult
\end{enumerate}

To free up some resources, the size of some \texttt{switch.p4} tables was reduced and some 
functionality was left out -- e.g., dataplane telemetry.
Rather than \texttt{ndp.p4}'s three trTCM meters per port
to track the output queue in all states, our \texttt{switch.p4}
implementation only uses two meters per port.

The first meter (\texttt{opti}) is used in the \textit{optimistic}
and \textit{halftimistic} states.
We set its Peak Information Rate (PIR) at line rate and the
Committed Information Rate (CIR) at half line rate.
When \texttt{opti} outputs red, this indicates that
traffic exceeds the line rate of
the port it was tracking, whereas
yellow indicates that traffic is greater than half
line rate but slower than line rate.
The second meter (\texttt{pesi}) has both rates set at one fourth line rate; 
its output is only used in \textit{pessimistic} state.
Based on the port state and the colors of the meters, we apply one of the actions
below:

{\small
\begin{table}[h!]
  \centering
  \begin{tabular}{|c | c | c | c |}
    \hline
    State & pessi color & opti color & Action  \\
    \hline\hline
    pessimistic & NOT GREEN & * & trim \\
    \hline
    halftimistic & * & NOT GREEN & trim \\
    \hline
    optimistic & * & RED & trim \\
    \hline
    pessimistic & GREEN & * & low prio queue \\
    \hline
    halftimistic & * & GREEN & low prio queue \\
    \hline
    optimistic & * & NOT RED & low prio queue \\
    \hline
  \end{tabular}
\end{table}}

To keep the state of a port, a ingress pipeline register stores the
timestamp of the last congestion signal.  The maximum size of a
register is 32 bits and there are few resources left in switch.p4, so
we can store only 32 bits of the pessimistic transition timestamp. A
Register Action is used to store the time of the pessimistic
transition and another Register Action is used to check how much time
has passed since the pessimistic transition.  Because of the limited
operations allowed in a Register Action, the case where the
timestamp overflows could not be included.

\lstset{
  basicstyle={\Small\ttfamily},
  stringstyle=\ttfamily,
  frame=single,
  emph={ON, if, else},
  emphstyle=\color{red},
  emph={[2]tstamp, last_pesi_trasition, state},
  emphstyle={[2]\color{blue}},
}
\begin{lstlisting}
if tstamp[31:0] > last_pesi_trasition + T1:
    last_pesi_trasition = 0
    state = OPTIMISTIC
else if (tstamp[31:0] > last_pesi_trasition + T0:
    state = HALFTIMISTIC
else:
    state = PESSIMISTIC
\end{lstlisting}

\section{Evaluation}

We have been using our Sonic implementation in a leaf-spine testbed
with four ToRs and three spines (Tofino switches), with 32 host NIC
ports of varying speeds attached in a balanced way to the ToRs. The switches
use BGP for routing, where each rack is advertised as a separate prefix.

We have run both high-throughput (storage) and short flow workloads
(microservices, memcached) in this setup, obtaining results that
closely match the ideal end-to-end behaviour of NDP\cite{ndp} and
EQDS\cite{eqds} stacks.  We include a subset of these results to this
paper for completeness; however the main goal of our evaluation is to
understand in detail the differences in the possible implementations
of trimming and the ideal behaviour as described in the NDP
paper\cite{ndp}, with the stated goal of showing that high quality
trimming implementations are possible on modern programmable hardware.

Our hardware experiments focus on our Tofino implementation only;
other groups have independently implemented 
trimming with on Broadcom Trident 4 and Nvidia Spectrum 2 switches
(via Mirror-on-Drop).

We rely on simulation to understand the merits of the alternative
ways of implementing trimming, and to validate the design and
parameter choices for our congestion feedback mechanism described in
\S\ref{sec:solution}.

To this end, we have implemented a model of our Tofino implementation
in the htsim simulator \cite{ndp}, as well as other mechanisms that
support trimming including Mirror-on-Drop, Deflect-on-Drop, ingress
meter trimming and so forth.  We use simulations to compare the
behaviour of a 64 port 100Gbps Tofino switch to that of an idealized
NDP switch having 64 ports at 100Gbps.

\subsection{Basic mechanism comparison}
\label{subsec:basic_mechanism_comparison}

\begin{figure*}
  \center
  \begin{subfigure}{0.33\textwidth}
    \includegraphics[width=\columnwidth]{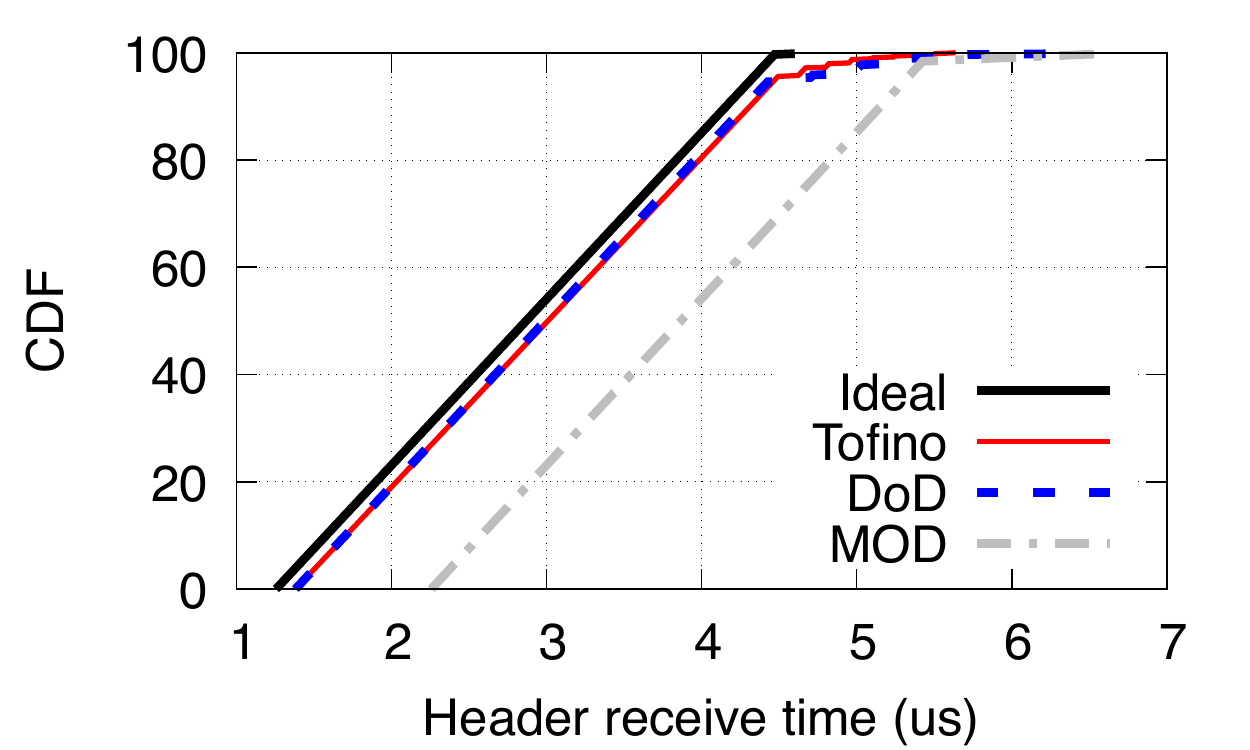}
    \caption{64 to 1 incast\label{fig:mechanism_comparison_a}}
  \end{subfigure}
  \begin{subfigure}{0.33\textwidth}
    \includegraphics[width=\columnwidth]{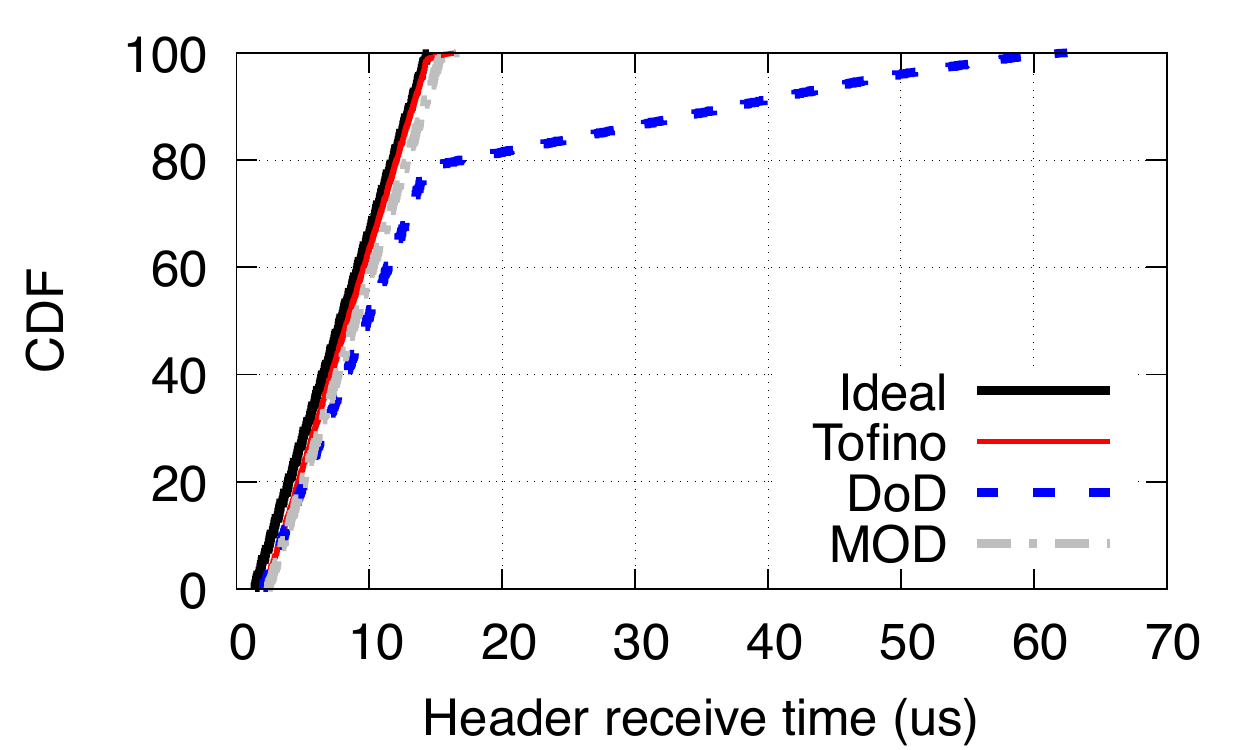}
    \caption{16 to 1 incast (4 times)\label{fig:mechanism_comparison_b}}
  \end{subfigure}
  \begin{subfigure}{0.33\textwidth}
    \includegraphics[width=\columnwidth]{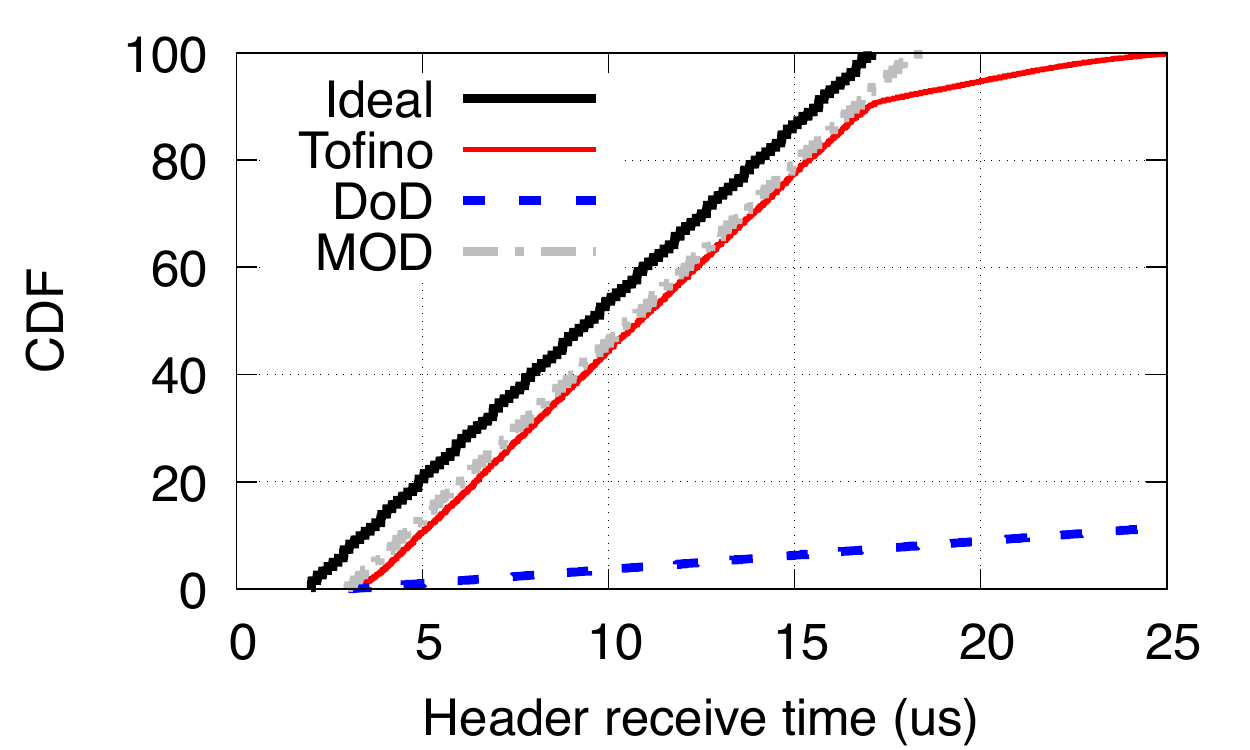}
    \caption{4 to 1 incast (16 times)\label{fig:mechanism_comparison_c}}
  \end{subfigure}
  \caption{Comparison between different ways of implementing trimming.}
  \label{fig:mechanism_comparison}
\end{figure*}

Which is the best way to implement trimming? To understand the tradeoffs, we compare the
idealized switch behaviour to:
\begin{itemize}
\item A generic Mirror-on-Drop (MoD) switch with a 100Gbps mirror-on-drop port per pipeline, and a recirculation latency of 1us (similar in spirit to Trident 4 and Spectrum implementations of trimming).
\item The Tofino with per-destination meters in the ingress pipeline and Deflect-on-Drop in case of queue overload.
\item Our full Tofino solution with pesi and half meters, as well as notifications from DoD to ingress pipelines.
\end{itemize}

We use simple incast workloads where all 64 ports send to one,
four, eight, or sixteen destination ports, and focus on the
uncontrolled startup phase where trimming is required, before receiver
pull-pacing kicks in. The results are shown in Figures
\ref{fig:mechanism_comparison}. The incast senders start at time 0 and
all figures show the CDF of header (trimmed packet) arrival times at
the receiver.

Figure \ref{fig:mechanism_comparison_a} shows header arrival times for
a 64 to 1 incast. The DoD-only implementation of Tofino works
similarly to our complete mechanism, because they both rely on ingress
pipeline trimming based on meters. Trimming ensures that each pipeline outputs 100Gbps
towards the target port, so a third of these packets are still deflected on drop; these
will experience higher latency due to the serialization and recirculation delay. As
the incast is short lived, the effects of these recirculated packets is small.

Mirror-on-Drop performs worst in this case as it always adds a delay
of around 1us to all headers because of the recirculation latency. For short-lived incasts,
retransmissions triggered by these headers will arrive 1us later and can lead to link under-utilization.
To avoid such cases, we need to add 1us (9 packets) extra buffering to each port, increasing buffering to
20 packets from the suggested 10 \cite{ndp}.

Figure \ref{fig:mechanism_comparison_c} shows another extreme where
the 64 senders send to 16 receivers, creating sixteen parallel 4-to-1
incasts.  This is the worst case scenario described in
Section~\ref{sec:trimming}, where the senders to the same receiver
are spread across different pipes. In this case, fixed ingress meters
(the ``DoD'' curve) do not reduce the load, and each pipe has an
excess of 1.2Tbps of traffic arriving that should be trimmed. This
overloads the 100Gb/s recirculation port and results in a huge latency
for these headers; this mechanism is unstable in this setup.

MoD works well in this case, its only downside being the 1us latency
we already discussed.  The congestion feedback we implemented helps
the Tofino behave similarly to MoD-capable switches; however, some
packets are still Deflected-on-Drop and the resulting headers
experience higher latency (90\% percentile and higher in the CDF).

Figure \ref{fig:mechanism_comparison_b} shows behaviour in between
these extremes: we have 16 to 1 incasts (4 in parallel), with 4
senders in each pipeline.  In this case, we have a lot of ingress
trimming and 300Gbps of excess traffic per pipe; the two modes in the
DoD curve show ingress trimming (first part) and DoD trimming (second,
lower slope). The adaptive ingress trimming in our full solution closely tracks the ideal. It is
slightly faster than MoD because of this ingress trimming;
the feedback ensures that few packets are sent through the DoD
mechanism, reducing latency.

In summary, Mirror-on-Drop works well but adds a fixed latency
which requires slightly larger buffers; trimming solutions based on
this have been implemented by other groups and tested on Broadcom
Trident 4 and NVidia Spectrum 2 switches \cite{eqds}, and we do not
explore this mechanism further here.

Ingress trimming is closest to ideal as long each meter sees
enough of the traffic load; Deflect-on-Drop adds large
latency to headers and can only cope with limited overload. Ingress trimming
coupled with our control loop tracks Ideal quite closely in
these experiments.

As this mechanism is quite involved, we focus the rest of this evaluation to show
that it is robust and works in all conditions, as well as explaining how we chose
its parameters.

\subsection{Base evaluation: Ideal versus Tofino}

\begin{figure*}
  \center
  \begin{subfigure}{.33\textwidth}
  \includegraphics[width=\columnwidth]{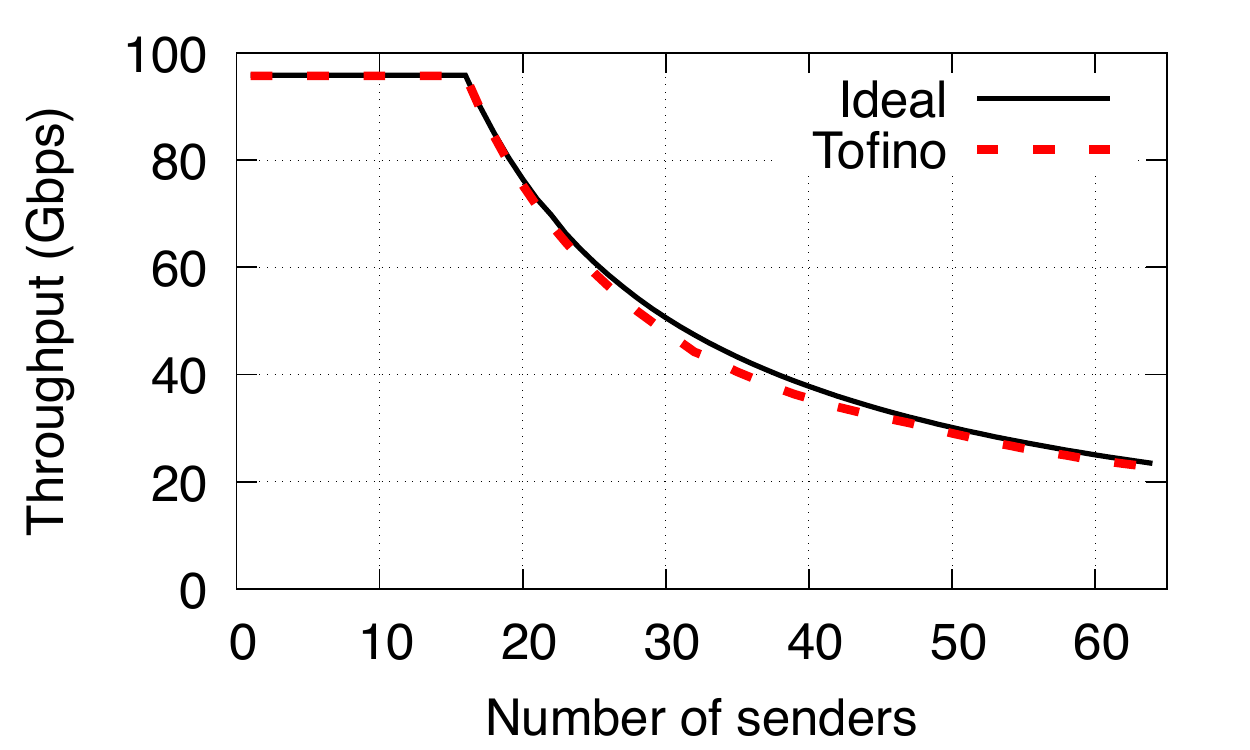}
  \caption{Average throughput.}
  \end{subfigure}
  \begin{subfigure}{0.33\textwidth}
    \includegraphics[width=\columnwidth]{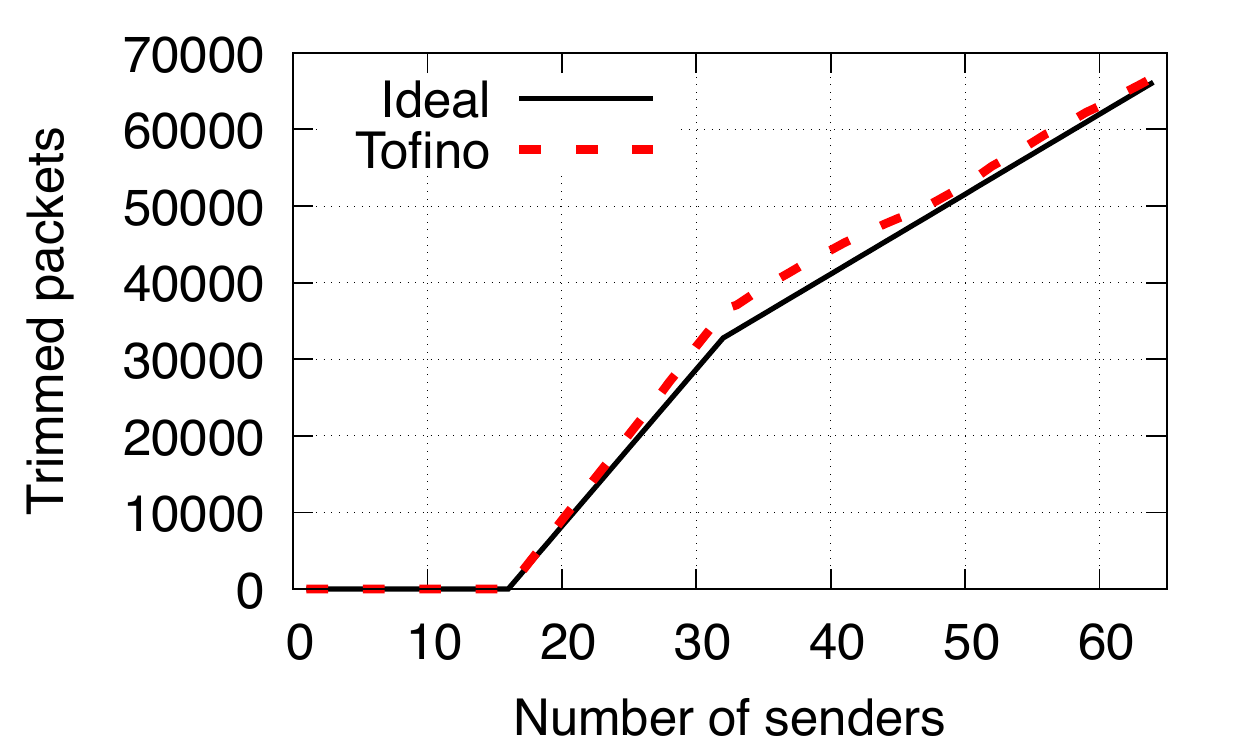}
  \caption{Number of trims.}    
  \end{subfigure}
  \begin{subfigure}{0.33\textwidth}
    \includegraphics[width=\columnwidth]{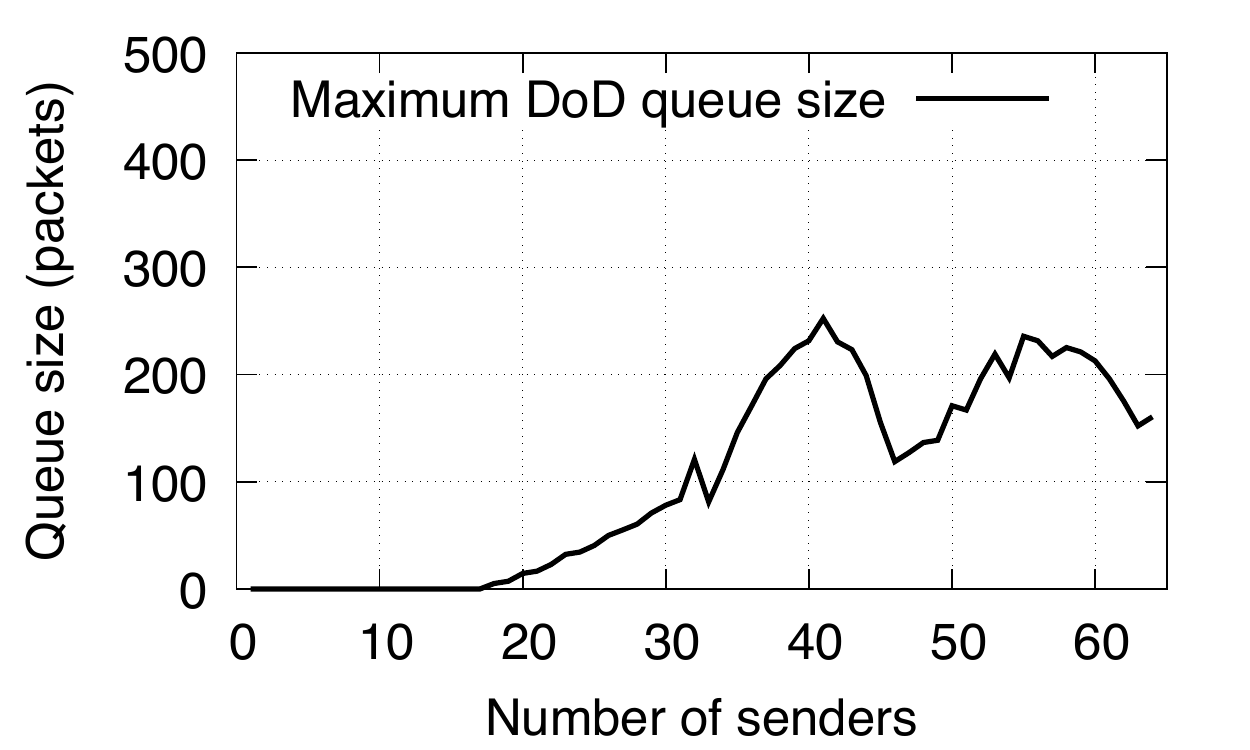}
    \caption{Maximum DoD queue size.}
  \end{subfigure}
  \caption{Comparison between Tofino and Ideal Switch implementations.}
  \label{fig:comparison}
\end{figure*}

To understand macroscopic behaviour of our Tofino implementation, we
now compare the average throughput obtained, the fairness, as well the
number of packets trimmed by each switch when we vary the number of
senders from 1 to 64. In each experiment,
sender $i$ connects the ingress port $i$ of the switch and sends traffic
to a receiver connected to the output port $i\%16$ of the switch. With
64 senders this corresponds to the worst case scenario for the Tofino
switch where each DoD queue has an overflow ratio of 12 to 1; however,
values smaller than 64 are also interesting to explore because they
create traffic imbalance across pipes where fairness could be affected by the wrong
congestion response mechanism.

Each sender opens one NDP connection to the receiver; when multiple
receivers are connected to the same port, they share the same pull
pacer which allows NDP to regulate the rate of incoming packets after
the initial sender-controlled part of the connection. This initial
phase is the most difficult one for the switch because pull pacing is
not in effect yet, and packets arrive at line rate from the senders;
to amplify its effects we use a large initial window for NDP of 1000
packets; this means that our senders will send at line rate for the
first ~125 microseconds; we run our experiment for 500 microseconds in
total.

When the switch dequeues a packet from the DoD queue for an outgoing port,
it sends congestion notifications to all pipelines at once,
instructing them to go into pessimistic mode for that port for 6
microseconds (the time to drain 64 packets). After that, the switch
goes into semi-pessimistic mode for another 18 microseconds.  The
latency of packet recirculation is set to 1us.

We plot the average throughput obtained by the flows in the experiment
in figure \ref{fig:comparison}.a. The goodput obtained by the Tofino
implementation of NDP is within 5\% of the ideal NDP behaviour.
Figure \ref{fig:comparison}.b shows that the Tofino trims at most 10\%
more packets than the ideal switch (6\% more on average across all
experiments). This is expected, since the congestion mechanism is
coarser grained than the ideal switch implementation where packets are
only trimmed when the queue is full. Still, the results below show
that our congestion mechanism manages to keep the DoD queues low;
across all experiments, the maximum DoD queue size did not exceed 250
packets.

To better understand the results, we now focus on a few specific
experiments, where we fix the number of senders and take a closer look
at how the congestion mechanisms come into effect.

The smallest number of senders that trigger a buildup of the DoD queue
is 18 senders where senders 1 to 16 send to output ports 1 to 16,
sender 17 to port 1 and 18 to 2. In this case, total traffic to ports
1 and 2 is 200Gbps each; 200Gbps from pipeline 1 and 200Gbps from
pipeline 2. A little less than half of this traffic enters the output
port, and half is directed to the DoD queue, trimmed and then makes it
to the output port. The arrival rate of each DoD queue in pipelines 1
and 2 is a little over 100Gbps, thus a queue builds and the congestion
mechanism kicks in.  Note that with 17 senders, the rate of traffic
arriving in the DoD queue is just 50Gbps, so a queue never builds, and
our congestion mechanism is not triggered; with less than 17 senders,
no traffic is dropped.

\begin{figure*}
  \begin{subfigure}{0.33\textwidth}
    \center
    \includegraphics[width=\columnwidth]{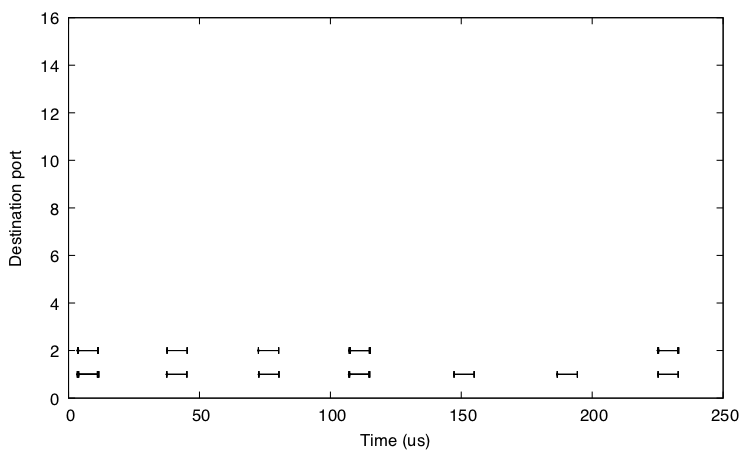}
    \caption{Congestion response for 18 senders.}
  \end{subfigure}
  \begin{subfigure}{0.33\textwidth}
    \center
    \includegraphics[width=\columnwidth]{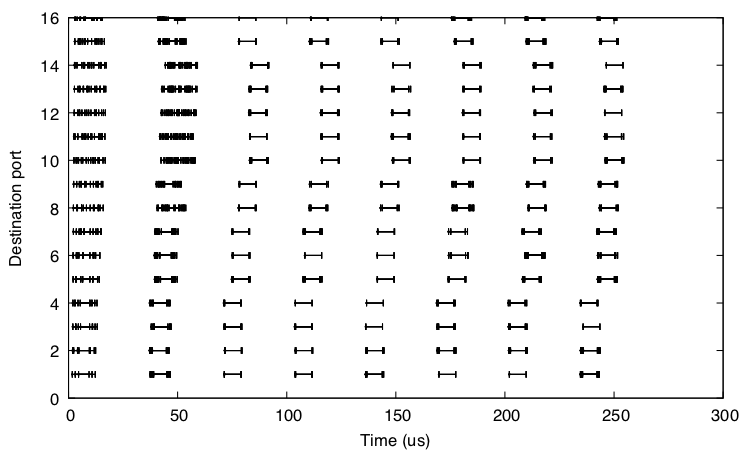}
    \caption{Congestion response for 32 senders}
  \end{subfigure}
  \begin{subfigure}{0.33\textwidth}
    \center
    \includegraphics[width=\columnwidth]{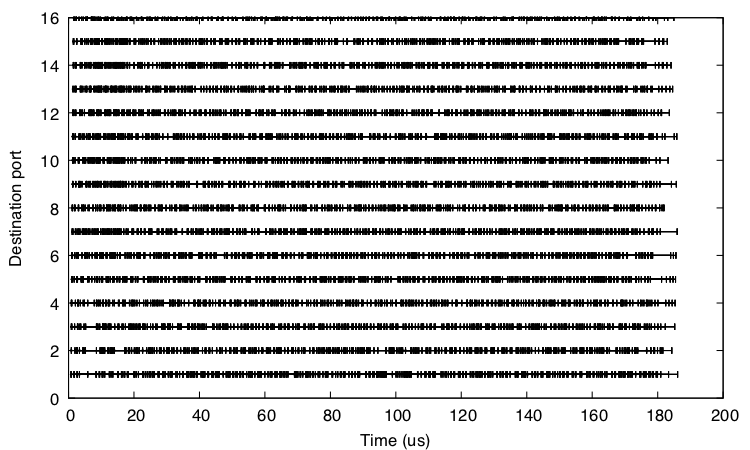}
    \caption{Congestion response for 64 senders}
  \end{subfigure}
  \caption{Congestion response: pessimistic mode activation, per port, as a function of time.}
  \label{fig:cong_response}
\end{figure*}

We plot the congestion mechanism at work in the 18 sender case
in Figure \ref{fig:cong_response}.a. The X axis shows time since the beginning of the experiment,
where 18 senders start blasting 1000 1.5KB packets each before waiting
for feedback from the receiver. The Y axis shows the destination port;
horizontal lines show periods where pessimistic mode is enabled for
the specified target port, on all pipelines. Note that, after a
pessimistic period ends, there is a halftimistic period
that kicks in that lasts for 24us (not shown).

The graph shows that congestion mode kicks in periodically for ports 1
and 2 (where congestion occurs); less than a fourth of the time
traffic is trimmed to 25Gbps, and most of the remainder it is trimmed
to half, and the remainder there is no trimming in the ingress
pipeline which triggers pessimistic mode again. The resulting flow
rates are 95.7Gbps for senders 3-16, and 46-47Gbps for senders 1,2,16
and 17; this is within 1\% of optimal behaviour.

We also plot below the congestion behaviour for 32 and 64 senders
respectively. For 32 senders, pessimistic mode is active for all ports
periodically, roughly a quarter of the time; half pessimistic mode for
the remainder; flow throughput is 45-46Gbps, near-optimal.


Finally, with 64 senders, the worst case from an overload point of
view, pessimistic trimming is active for the entire duration of the
incast, and flow bandwidth are fairly distributed (22.2Gbps to
23.4Gbps) and within 1\% of the optimal.

\begin{figure*}
  \center
  \begin{subfigure}{0.33\textwidth}
    \includegraphics[width=\columnwidth]{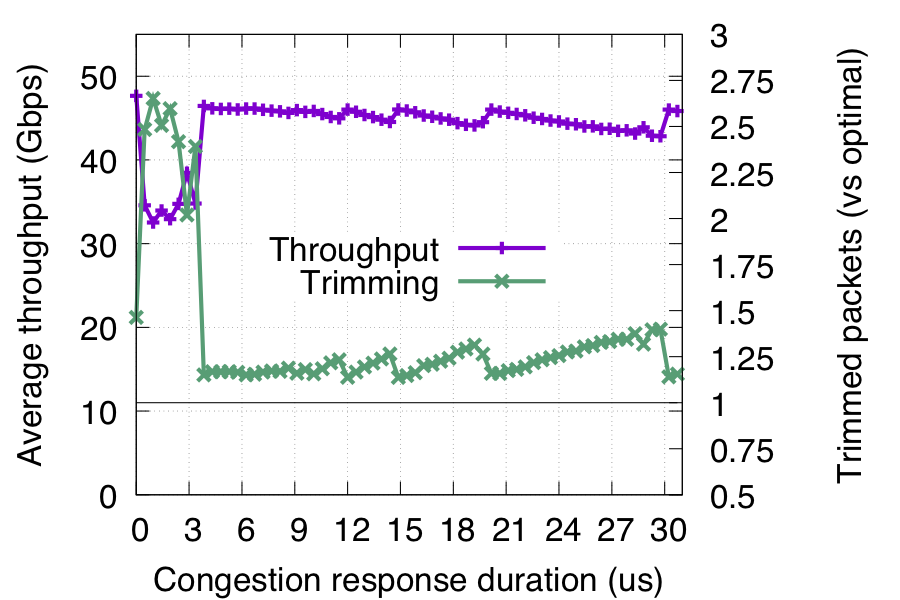}
    \caption{32 senders}
  \end{subfigure}
  \begin{subfigure}{0.33\textwidth}
    \includegraphics[width=\columnwidth]{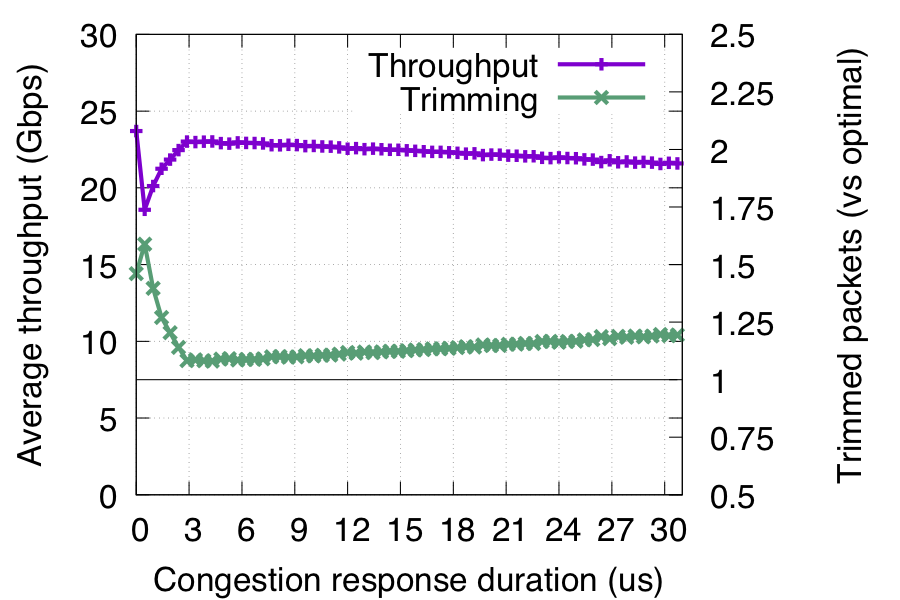}
    \caption{64 senders}
  \end{subfigure}
  \begin{subfigure}{0.33\textwidth}
    \center
    \includegraphics[width=\columnwidth]{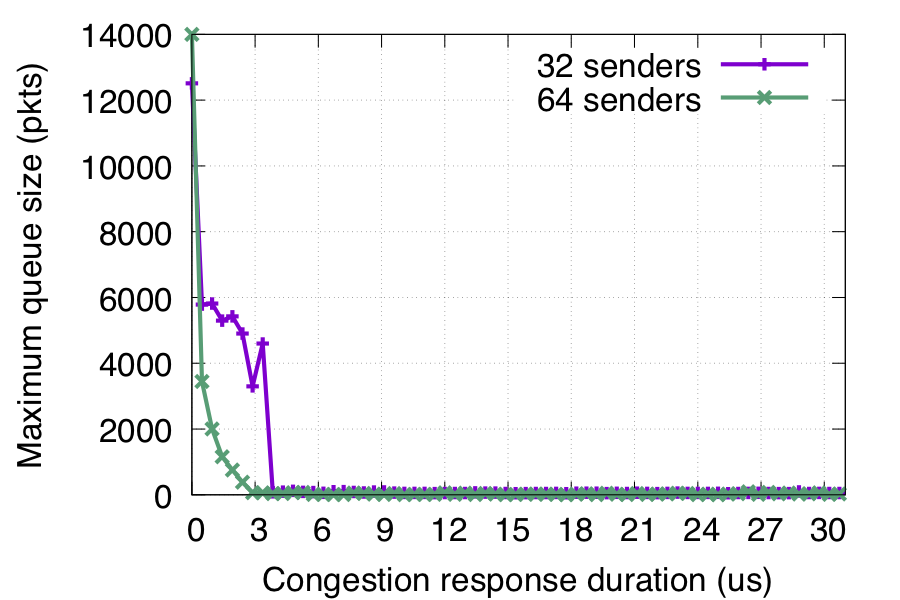}
    \caption{DoD queue size}
  \end{subfigure}
  \caption{Effect of congestion response duration on switch behaviour}
  \label{fig:response-duration}
\end{figure*}

\subsection{Sensitivity analysis}

The experiments so far were run with 6us pessimistic mode followed by
18us halftimistic mode. How do results vary when changing the
congestion response duration? More generally, how do each of our design
decisions affect the trimming behaviour. We answer these questions here.

First, we vary the congestion response from 0us (no response) to 30us
(the time needed to serialize 250 packets of 1500B) and examine the
number of trimmed packets compared to an ideal implementation, the
average throughput obtained as well as the maximum DoD queue size,
shown in Figure \ref{fig:response-duration}.c.

The maximum DoD queue size grows to almost 14000 packets when no
congestion response occurs (the point for 0us on the X axis in the
plot); this is because in each pipeline the first 16000 packets arrive
at line rate (1000 per ingress port), and roughly a quarter will make
it to the egress queue, while the others are deflected to the DoD
queue corresponding to the ingress pipeline.  To enable the DoD queue
to drain by X packets, it should receive no packets for the duration
it takes to serialize X packets at 100Gbps. Thus, all the ingress
pipelines should go into pessimistic mode for a duration at least as
large.

In the worst case scenario with 64 senders, in each RTT 48 packets
will be deflected on drop, roughly 12 to each DoD queue: 1.44us is
therefore the minimum duration for the congestion response to have an
effect. Experiments show that at least 3us ($\sim$30 packets time) are
needed to control the DoD queue size properly; this is because packets
for different ports may be ordered randomly in the DoD queue, and the
congestion response may focus only on one output port for a short
period of time, while allowing all other output ports to pump packets
in the DoD queue. In other words, the congestion response should be
long enough to ensure that it bridges the gaps in the DoD queue
between packets for the same output port.

The queue size results show that any congestion duration above 3us
manages to keep the queue size small; how about throughput and trimmed packets? The
results are shown in Figures \ref{fig:response-duration}.a-b. First, note that disabling the
congestion response means that the average throughput is near optimal,
at the cost of delayed congestion information and inflated short flow
completion times. When the congestion response is enabled but its
duration is less than 3us, the DoD queue size grows large and the
congestion response kicks in eventually; in this case, the switch ends
up trimming a lot more packets than it should, hurting
throughput. Beyond 3us, there is a wide range of values that work
well. However, the longer the congestion response, the bigger the
effect on throughput and other traffic following the incast. That is
why we chose 6us for our implementation. With these values, the number
of trimmed packets is just 10\% more than optimal.

\headline{Do we need a half-timistic mode?}
In our initial design, we only used a single 25\% pessimistic
response to congestion; this worked well for the worst case scenario
with 64 senders, but less so when congestion was less severe, for
instance when there were only 32 senders. In such cases, the on/off
 congestion response (fundamental to the experiment,
which was using 25\% trimming periodically to reach a 50\% trimming
average) resulted in DoD queue oscillations and the need for a large
congestion response duration to ensure stability.

\begin{figure*}
  \center
  \begin{subfigure}{0.33\textwidth}
    \includegraphics[width=0.9\columnwidth]{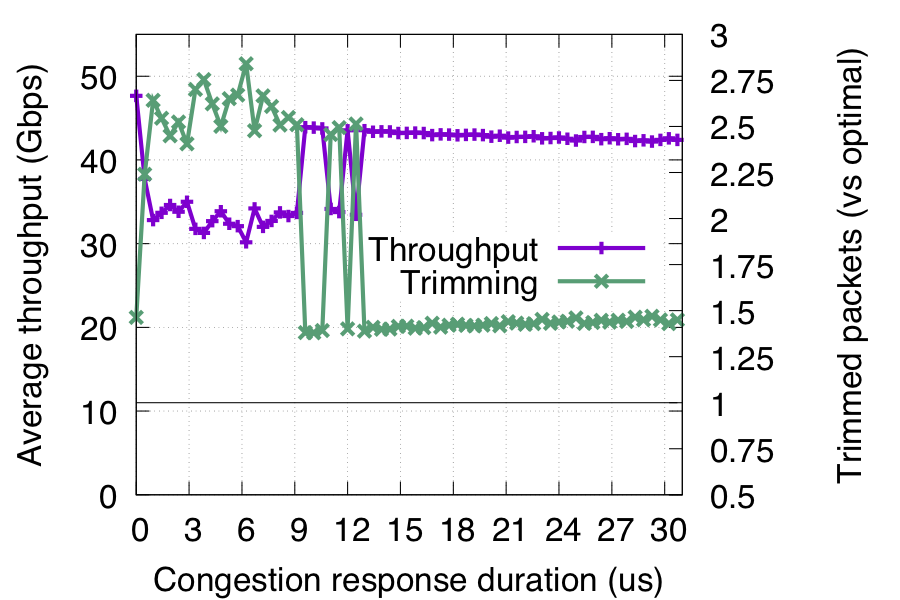}
    \caption{25\% trimming.}
  \end{subfigure}
  \begin{subfigure}{0.33\textwidth}
    \includegraphics[width=0.9\columnwidth]{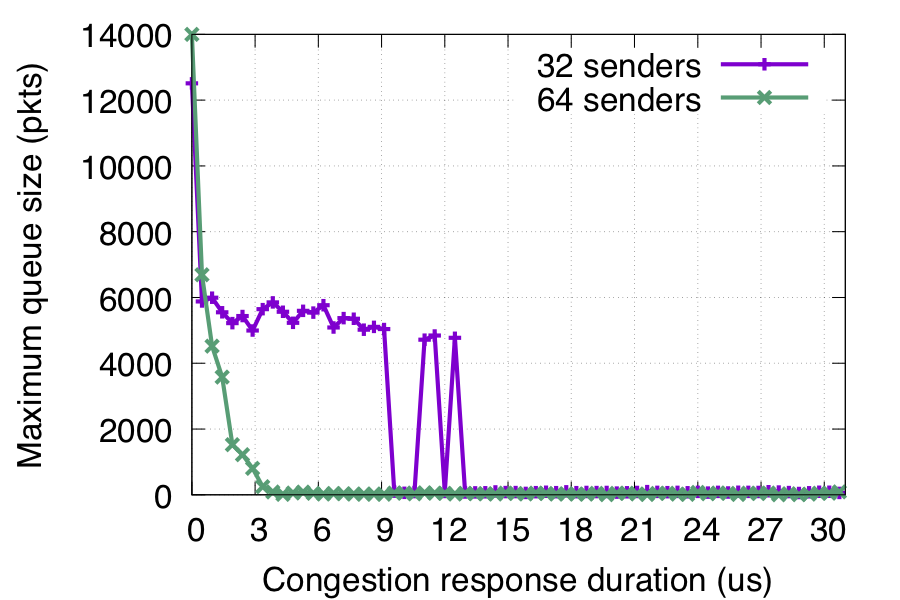}
    \caption{Queue: 25\% trimming.}
  \end{subfigure}
  \begin{subfigure}{0.33\textwidth}
    \includegraphics[width=0.9\columnwidth]{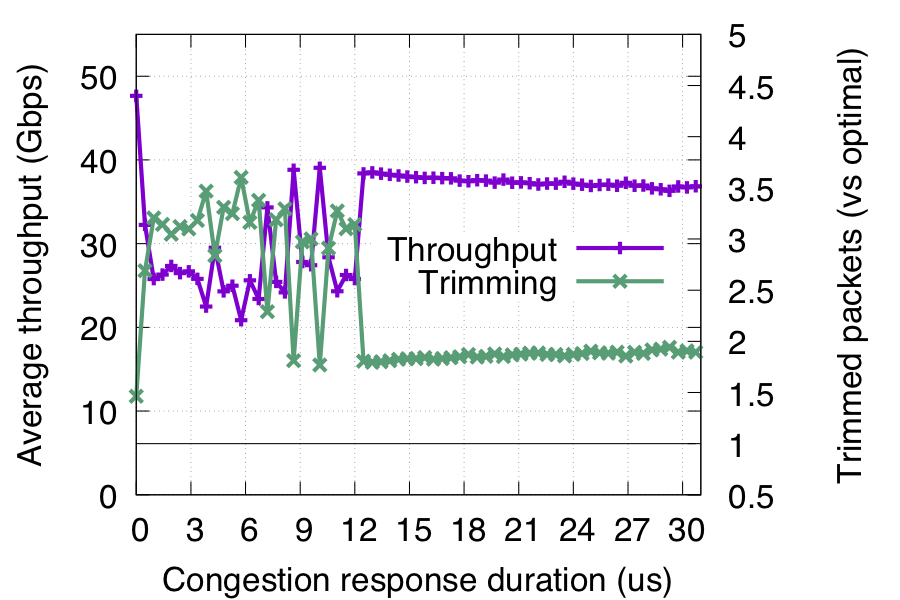}
    \caption{100\% trimming.}
  \end{subfigure}
  \caption{Queue oscillations appear with 32 senders with more aggressive trimming}
  \label{fig:pesi}
\end{figure*}

Figure \ref{fig:pesi} shows what happens with alternative trimming approaches, with 32 senders.
In \ref{fig:pesi}.a we use pessimistic mode alone, without a halftimistic mode following
it: this means that the ingress pipeline goes from 25\% trimming to no
trimming directly, instead of first trimming at 50\% for a period of
time. First, note that the results for 64 senders are quite similar (not shown),
since pessimistic mode is active for the duration of the experiment,
and a congestion response of 3us suffices also in this setup.

For 32 senders, the results are quite different: the DoD queue is
unstable until the congestion response duration exceeds 14us, and this
results in an average throughput of 42Gbps compared to 46Gbps for the
case when half pessimistic mode is used too.


Why not trim all
packets, then, instead of a fraction of packets?  Trimming all packets has
the benefit of allowing the DoD queue AND the output port queue to
drain, so it is a tougher congestion response. We run the
same experiment and plot the results in Figure \ref{fig:pesi}.c. 
The results show at least 12us
are required to have stable behaviour, while the trimming rate is 80\%
higher than optimal, compared to 15\% in our final
implementation. Throughput also suffers, with a maximum throughput of
38Gbps compared to 46Gbps. With 64 senders, the minimum response that
gives stable behaviours is 10us, and the average throughput is 19Gbps
compared to 23Gbps in our solution.

\headline{Why not inform a single pipeline?}
In our final what-if scenario, we explore what happens when the
congestion signal is sent only to the pipeline that originated the DoD
packet. The congestion response is pessimistic mode followed by
half-pessimistic mode, but only for that pipeline.
The results show that, for both 32 and 64 senders, we now need a congestion
response duration of at least 9-12us for the DoD queues to remain
small; nevertheless, the number of trimmed packets is 10\%-15\% larger
than in our solution, and 20-25\% more than the ideal implementation.
  

\begin{figure*}
  \center
  \begin{minipage}{.33\textwidth}
    \center
    \includegraphics[width=\columnwidth]{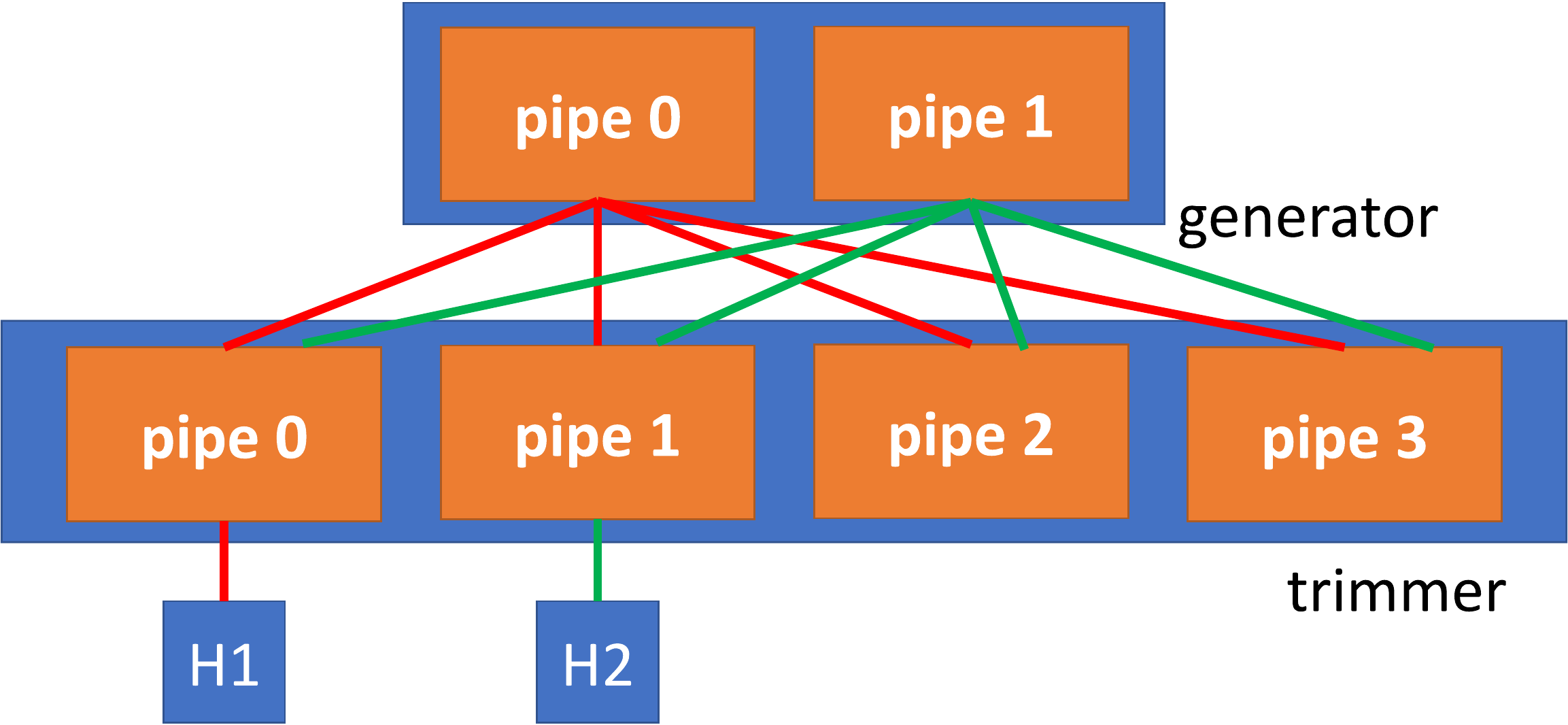}
    \caption{Tofino validation experimental setup}
    \label{fig:dodexp}
  \end{minipage}
  \begin{minipage}{.66\textwidth}
  \begin{subfigure}{0.49\textwidth}
  	\includegraphics[width=\columnwidth]{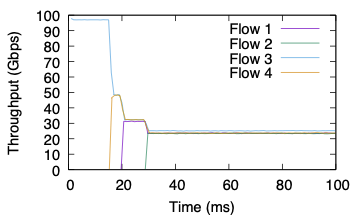}
  	\caption{Tofino ndp.p4 implementation}
  	\label{fig:tofino-throughput}
  \end{subfigure}
  \begin{subfigure}{0.49\textwidth}
  	\includegraphics[width=\columnwidth]{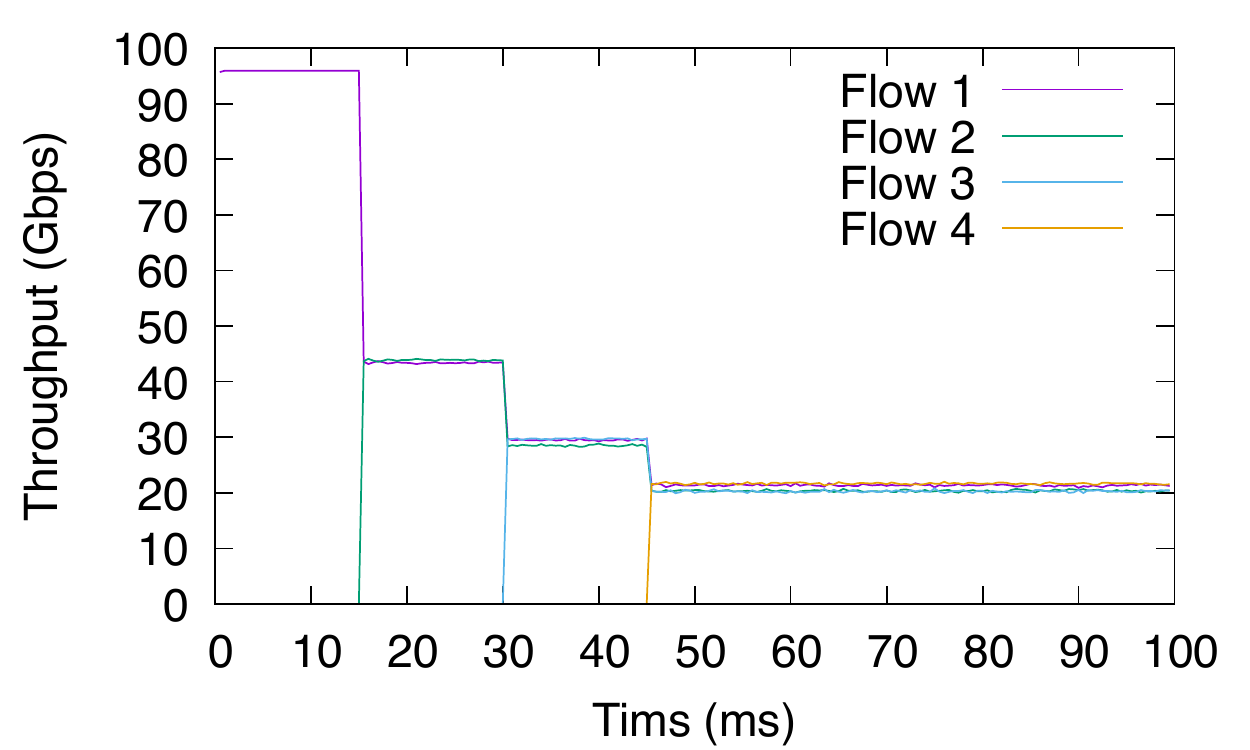}
  	\caption{Simulated Tofino algorithm}
	\label{fig:sim-throughput}
  \end{subfigure}
  \caption{Throughput of the Tofino implementation with DoD}
  \label{fig:tofino-throughput}
  \end{minipage}
\end{figure*}

\subsection{Tofino evaluation}

Most of the evaluation in this paper uses simulation.  Simulation
allows us to precisely start flows at exact times, compare different
algorithms with idealized trimming, and perform fine-grain
instrumentation of our feedback loops.  In the end, even though
the algorithm is the same, what we really want to see is how trimming
performs in real hardware.

To generate controlled 100Gb/s flows we use a second Tofino switch
(``generator'') to generate flows, as in Figure \ref{fig:dodexp}.
The {\texttt trimmer} switch runs our ndp.p4 implementation.  Four
flows shown in red are sent from the generator to host H1 and four
shown in green go to H2, resulting in a persistent high trimming rate.
Such trimming would be transient with an end-to-end control loop such
as NDP that kicks in after the first RTT, but the purpose here is to
stress-test the switch.  We verified that in this scenario no packets
are lost - every packet is either delivered completely or its header
arrives at the receiver.

Figure~\ref{fig:tofino-throughput} shows the throughput of the four
flows arriving at H1 during the startup phase as each flow arrives.
The sharing is almost perfect with no unexpected startup transients.
For comparison, ~\ref{fig:sim-throughput} shows the same setup running
in our simulator, demonstrating that the simulated behavior closely
matches that of the real hardware.

\begin{figure*}
  \begin{minipage}{0.33\textwidth}
	\center
	\includegraphics[width=\columnwidth]{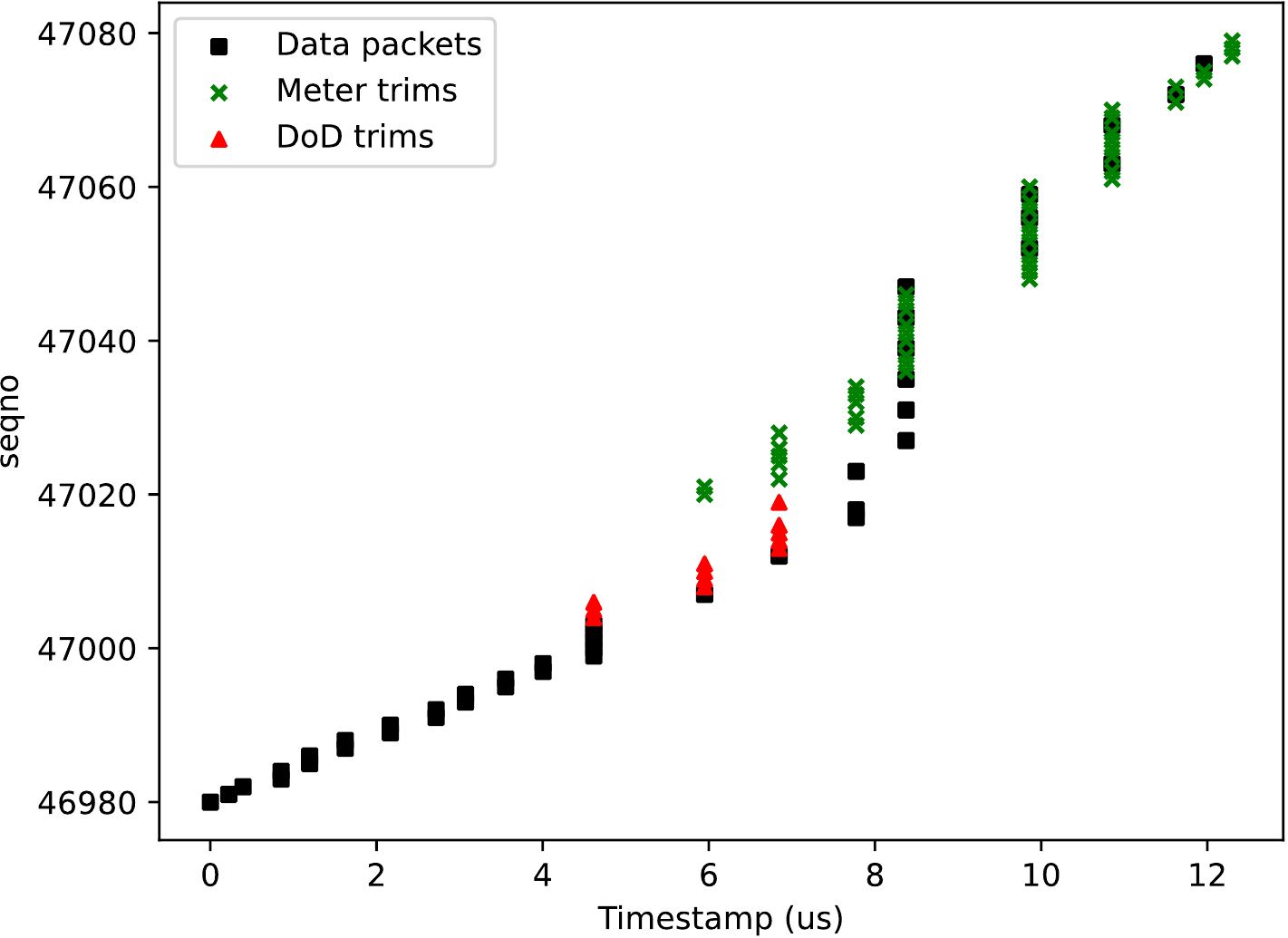}
	\caption{Trimming in Tofino}
	\label{fig:tofino_2_1}
\end{minipage}
\begin{minipage}{0.33\textwidth}
\center
    \includegraphics[width=\columnwidth]{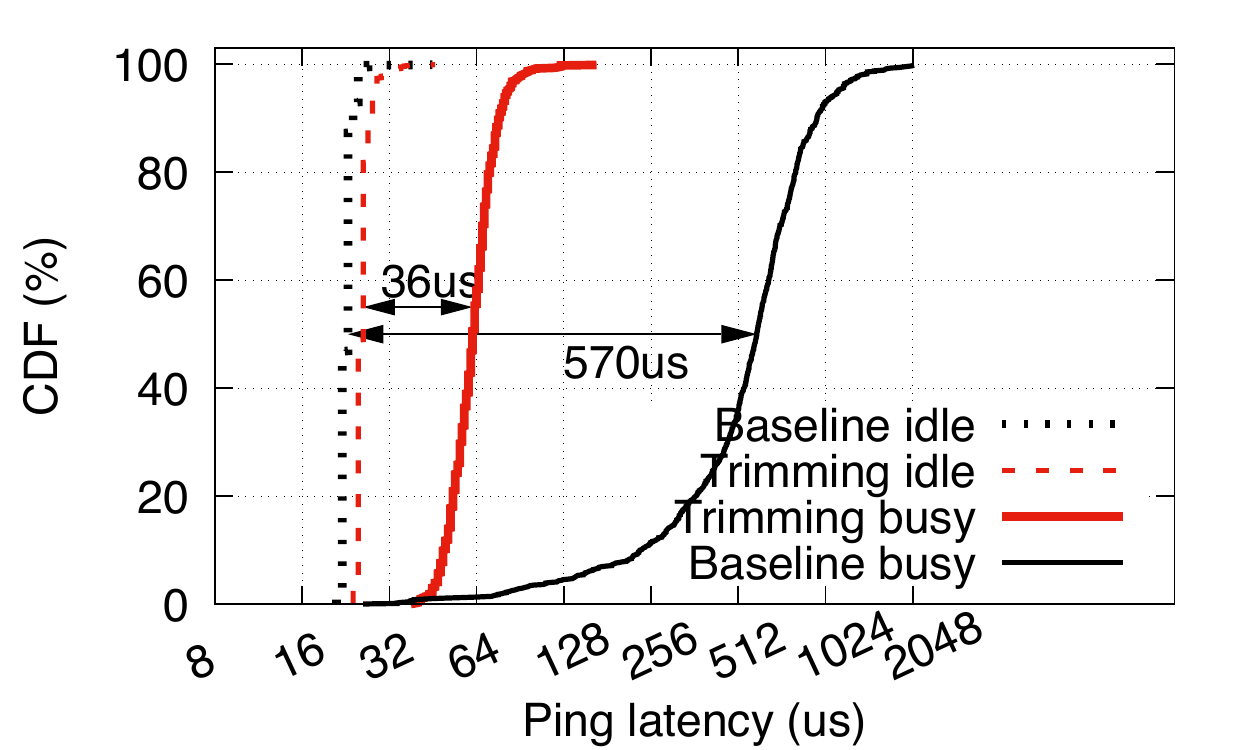}
    \caption{Faster pings with trimming}
    \label{fig:latency}
\end{minipage}
\begin{minipage}{0.33\textwidth}
\center
    \includegraphics[width=\columnwidth]{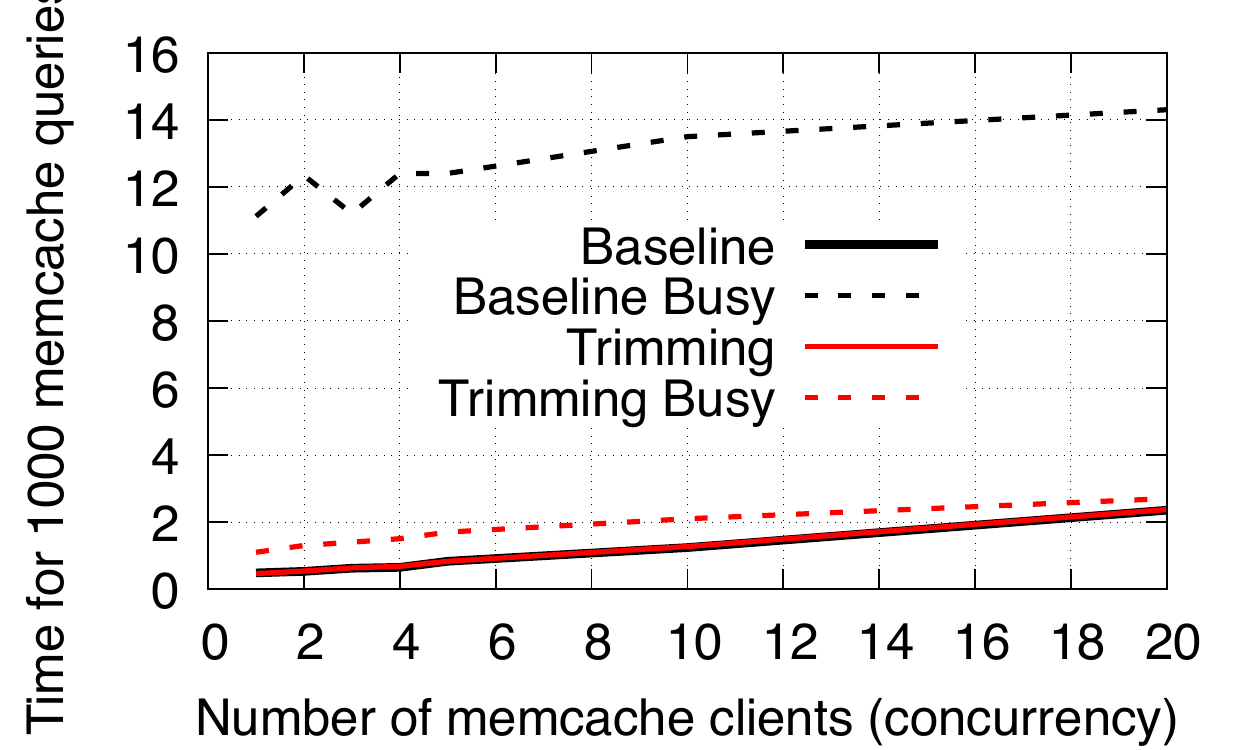}
    \caption{Memcached with trimming}
    \label{fig:memcached}
\end{minipage}
\end{figure*}

We modified our ndp.p4 implementation to mark packets that were
ingress trimmed differently from those that were DoD-trimmed. 
We record traces of 20k packets during a 4:1 incast
resulting both from the simulator and from the harwdware Tofino switch.
The table below shows the same number of trims happens, though more of
them were DoD trims in the simulator than in the Tofino.
This difference is likely due to differences in flow startup timing.  As
expected, most trims happen in the ingress trimmer.

\begin{table}[h!]
	\begin{tabular}{|l|l|l|}
		\hline
		Implementation & Ingress trims & DoD trims \\ \hline
		Simulator & 13.49K & 682 \\ \hline
		Tofino & 13.5k & 678 \\ \hline
	\end{tabular}
\end{table}

Figure~\ref{fig:tofino_2_1} shows packets from the first flow arriving
at H1 when the second flow to H1 starts at t=0.  The figure plots
sequence number against measured arrival time at H1. The arrival
times are rather bursty due to DPDK batching at H1, which is needed to
keep up with the arriving 100Gb/s stream.  From t=0 until 4.7$\mu s$ we can
see data packets arriving, but the gradient is below the trend line as
a queue is building in the switch. At 4.7$\mu s$, the queue overflows and a
number of packets enter the DoD pipeline. The congestion feedback then
kicks in, and at 6$\mu s$ the first ingress-trimmed packet arrives.  All
the early ingress-trimmed packets jump the data packet queue in the
switch, arriving significantly ahead of untrimmed data packets sent
around the same time.  These low latency trimmed packets are very
useful for NDP's control loop.  Finally, by 10$\mu s$, the switch queue has
largely drained and the network latency is very low again -- half the
packets are trimmed (the demand is 200Gb/s into a 100Gb/s port), but
there is no significant queue left to jump, so they arrive with
similar latency to untrimmed data packets.





\subsection{End-to-end experiments}

To round up our evaluation, we now present results from application
workloads we have run on our leaf-spine testbed that highlight the
benefits of packet trimming over conventional tail-drop or ECN-based
switches. All the links in our testbed are set to 25Gbps to ensure we
do not measure host overheads which would pollute the results at
100Gbps speeds. We compare the vanilla setup, with kernel Linux hosts and
tail-drop switches, to the Tofino setup where trimming is implemented in all switches,
and our hosts run the EQDS stack \cite{eqds}.

The key advantage of trimming is its ability to use small packet
buffers while allowing host stacks to react quickly to overload and to
cope with reordering.

The bisection bandwidth of our network is 75Gbps (three spines) and as
a first test we run three parallel connections between two racks. In
the vanilla setup, frequently flows are mapped by ECMP onto the same
spine, halving the achieved bandwidth.  With trimming, the end-hosts
can do packet-level load balancing and all flows reach 25Gbps
constantly.

In our next experiments, we explore the benefits of small switch
buffers to latency-sensitive applications. Our first experiment is simple:
we ping a destination machine
when it is idle, and when it has nine other TCP connections sending to
it from separate machines, plotting the measured ping-times in Figure
\ref{fig:latency}. Note that the baseline implementation, despite having only
a 200-packet buffer (fairly low for TCP standards), adds a 1/2ms delay to the
latency sensitive traffic when the network is busy. With trimming, we
can configure the switch buffer to be much lower (just 12 packets), which
causes a much smaller latency increase of just 36us.

Finally, we install \texttt{memcached} on one target machine and use a
benchmarking tool called \texttt{memcslap} to test its performance;
the benchmark consists of issues 1000 memcache requests in a closed
loop for each emulated client. We run the experiment both when the
target machine is idle, and when it is busy with a long lived TCP
incast which will fill the ToR switch buffer corresponding to the
destination host.

Figure \ref{fig:memcached} shows the total time needed to run the
\texttt{memcslap} benchmark; when the destination is idel, it takes
between one to two seconds depending on the number of emulated
clients.  When the destination is busy, however, the baseline time
jumps more than ten times; in contrast, trimming enables keeping queue
sizes small and the increase is much smaller.

\section{Conclusions}

Although current programmable switches were not specifically designed
with packet trimming in mind, they are sufficiently flexible that it
is now possible to implement and deploy trimming in today's
datacenters that deploy such switches.  The main benefits of trimming
occur when implemented in top-of-rack switches which are most likely
to support such programmability. 

Key to enabling trimming is the
ability to deflect packets that would have been dropped
into a recirculation port.  On Tofino we have demonstrated that even
though the deflect-on-drop bandwidth is limited, combining this with
P4-based trimming in the ingress pipelines using meters enables a
low-latency trimming solution that closely mirrors idealized trimming.
Switches such as Tofino 2 that allow the ingress pipeline to read the
downstream queue size, or switches that allow mirroring just the
header to a recirculation pipeline can also effectively implement
trimming.

For minimum feedback latency, ideally most trimming will occur in the
ingress pipeline without the headers being recirculated, as in our
Tofino implementation.  When some trimming is done via recirculation
and some in ingress, header reordering will occur.  The great
advantage of trimming, though, is that reordering need not be mistaken
for loss. Indeed, protocols such as NDP or EQDS designed with packet
spraying and trimming in mind have no problem with such reordering.

\printbibliography
\end{document}